\documentclass[aps,prd,onecolumn,showpacs,showkeys,amsmath,amssymb]{revtex4}
\usepackage{amsfonts}
\usepackage{amsmath}
\usepackage{graphicx}
\usepackage{subfig}
\usepackage{dcolumn}
\usepackage{natbib}
\usepackage{bm}
\usepackage{booktabs}
\usepackage{slashed}

\usepackage{ulem}
\usepackage{color}

\makeatletter
\newcommand{\figcaption}{\def\@captype{figure}\caption}
\newcommand{\tabcaption}{\def\@captype{table}\caption}

\newcommand{\Rmnum}[1]{\expandafter\@slowromancap\romannumeral #1@}
\def\hlinewd#1{%
  \noalign{\ifnum0=`}\fi\hrule \@height #1 \futurelet
   \reserved@a\@xhline}
\makeatother
\def\dab{\int^{\alpha_{max}}_{\alpha_{min}}d\alpha\int^{\beta_{max}}_{\beta_{min}}d\beta}
\def\qq{\langle\bar qq\rangle}
\def\ss{\langle \bar ss\rangle}
\def\GGa{\langle GG\rangle}
\def\GGb{\langle g_s^2GG\rangle}
\def\qGqa{\langle\bar qg_s\sigma\cdot Gq\rangle}
\def\qGqb{\langle\bar qGq\rangle}
\def\sGsa{\langle\bar sg_s\sigma\cdot Gs\rangle}
\def\sGsb{\langle\bar sGs\rangle}
\def\f(s){(\alpha m_s^2+\beta m_Q^2-\alpha\beta s)}
\def\non{\\ \nonumber}

\begin{document}
%
%
\title{Open-flavor charmed/bottom $sq\bar q\bar Q$ and $qq\bar q\bar Q$ tetraquark states}

 \author{Wei Chen$^1$}
 \author{Hua-Xing Chen$^2$}
 \email{hxchen@buaa.edu.cn}
 \author{Xiang Liu$^{3,4}$}
 \email{xiangliu@lzu.edu.cn}
 \author{T. G. Steele$^1$}
 \email{tom.steele@usask.ca}
 \author{Shi-Lin Zhu$^{5,6,7}$}
 \email{zhusl@pku.edu.cn}
 \affiliation{
 $^1$Department of Physics and Engineering Physics, University of Saskatchewan, Saskatoon, Saskatchewan, S7N 5E2, Canada
 \\
 $^2$School of Physics and Beijing Key Laboratory of Advanced Nuclear Materials and Physics, Beihang University, Beijing 100191, China
 \\
 $^3$School of Physical Science and Technology, Lanzhou University, Lanzhou 730000, China
 \\
 $^4$Research Center for Hadron and CSR Physics, Lanzhou University and Institute of Modern Physics of CAS, Lanzhou 730000, China
 \\
 $^5$School of Physics and State Key Laboratory of Nuclear Physics and Technology, Peking University, Beijing 100871, China
 \\
 $^6$Collaborative Innovation Center of Quantum Matter, Beijing 100871, China
 \\
 $^7$Center of High Energy Physics, Peking University, Beijing 100871, China
 }

\begin{abstract}
We provide comprehensive investigations for the mass spectrum of exotic open-flavor charmed/bottom $sq\bar q\bar c$,
$qq\bar q\bar c$, $sq\bar q\bar b$, $qq\bar q\bar b$ tetraquark states with various spin-parity assignments
$J^P=0^+, 1^+, 2^+$ and $0^-, 1^-$ in the framework of QCD sum rules. In the diquark configuration,
we construct the diquark-antidiquark interpolating tetraquark currents using the color-antisymmetric scalar
and axial-vector diquark fields. The stable mass sum rules are established in reasonable parameter
working ranges, which are used to give reliable mass predictions for these tetraquark states.
We obtain the mass spectra for the open-flavor charmed/bottom $sq\bar q\bar c$, $qq\bar q\bar c$,
$sq\bar q\bar b$, $qq\bar q\bar b$ tetraquark states with various spin-parity quantum numbers.
In addition, we suggest searching for exotic doubly-charged tetraquarks, such as $[sd][\bar u\bar c]\to
D_s^{(\ast)-}\pi^-$ in future experiments at facilities such as BESIII, BelleII, PANDA, LHCb and CMS, etc.
\end{abstract}


 \maketitle
%
\section{Introduction}\label{Sec:Intro}
During the past 14 years, there are many unexpected hadrons observed experimentally, such as the XYZ states~\cite{2016-Patrignani-p100001-100001} and hidden-charm pentaquarks \cite{2015-Aaij-p72001-72001}, etc. These resonances cannot be interpreted as
conventional quark-antiquark mesons or three-quark baryons in the conventional quark model \cite{2007-Klempt-p1-202}. They are exotic hadron candidates,
whose significant experimental and theoretical progress have been reviewed in Refs.~\cite{2016-Chen-p1-121,2015-Esposito-p1530002-1530002,2015-Olsen-p101401-101401,2016-Lebed-p-}.

Very recently, the D0 Collaboration reported the evidence for the narrow structure $X(5568)$ in the $B_s^0\pi^\pm$
invariant mass spectrum with 5.1$\sigma$ significance \cite{2016-Abazov-p22003-22003}. Its mass
and decay width were measured to be $m_{X(5568)}=5567.8\pm2.9(\rm stat)^{+0.9}_{-1.9}(\rm syst){\rm~MeV}$ and
$\Gamma_{X(5568)}=21.9\pm6.4(\rm stat)^{+5.0}_{-2.5}(\rm syst){\rm~MeV}$, and its spin-parity quantum number
was determined to be either $J^P=0^+$ or $1^+$. Later, the LHCb and CMS collaborations also
performed their analyses of the $pp$ collision data at energies $\sqrt{s}=7$ TeV and 8 TeV to search for the
$X(5568)$ state \cite{2016-Aaij-p152003-152003,2016-Collaboration-p-a}, but they did not find any unexpected
structure in the $B_s^0\pi^\pm$ invariant mass distribution. However, the D0 Collaboration themselves confirmed the
$X(5568)$ meson in the $B_s^0\pi^\pm$ invariant mass distribution with another channel $B_s^0\to D_s\mu\nu$
at the same mass and at the expected width and rate \cite{D0:X5568}.

Reported in the $B_s^0\pi^\pm$ final states, the $X(5568)$ meson, if it exists, could be a bottom-strange
$su\bar{d}\bar{b}$ (or $sd\bar{u}\bar{b}$) tetraquark state with valence quarks of four different flavors. To date,
the $X(5568)$ resonance has trigged many theoretical studies, including the diquark-antidiquark tetraquark
state~\cite{2016-Chen-p22002-22002,2016-Agaev-p74024-74024,2016-Zanetti-p96011-96011,2016-Wang-p335-339,2016-Wang-p279-279,2016-Wang-p93101-93101,2016-Tang-p558-558,2016-Agaev-p114007-114007,2016-Dias-p235-238,2016-Albuquerque-p1650093-1650093,2016-Liu-p74023-74023,2016-Stancu-p105001-105001,2016-Ali-p34036-34036,2016-He-p92-97,2016-Burns-p627-633,2016-Guo-p593-595,2016-Lue-p-,2016-Goerke-p-,2016-Agamaliev-p-}, hadron molecule~\cite{2016-Agaev-p351-351,2016-Kang-p54010-54010,2016-Chen-p351-351,2016-Albaladejo-p515-519,2016-Lang-p74509-74509,2016-Chen-p34006-34006,2016-Lu-p-,2016-Sun-p-},
non-resonant schemes \cite{2016-Liu-p455-455}, hybridized tetraquark model \cite{2016-Esposito-p292-295}, and so on.
One can find the detailed introduction for these theoretical studies in the recent review paper~\cite{2016-Chen-p-}.

In the charm sector, the two narrow charm-strange mesons $D_{s0}^\ast(2317)$ and $D_{s1}(2460)$ were
observed in 2003 in the $D_s^+\pi^0$ and $D_s^{\ast+}\pi^0$ invariant mass distributions by the BaBar
\cite{2003-Aubert-p242001-242001} and CLEO \cite{2003-Besson-p32002-32002} collaborations, respectively.
Their observed masses are $(2317.7\pm0.6)$ MeV
and $(2459.5\pm0.6)$ MeV, respectively, which are much lower than the Godfrey–Isgur (GI) model predictions
\cite{1991-Godfrey-p1679-1687}. These states quickly attracted many theoretical studies involving various exotic
assignments \cite{2003-Barnes-p54006-54006,2004-Chen-p232001-232001}, which can also be found in the
review paper Ref.~\cite{2016-Chen-p-} and its related
references. Among these configurations, the four-quark assignment $cq\bar s\bar q$ is particularly interesting.

Inspired by the experimental information and theoretical studies of the $X(5568)$, $D_{s0}^\ast(2317)$ and
$D_{s1}(2460)$
mesons, we shall provide comprehensive studies for the open-flavor charmed/bottom $sq\bar q\bar Q$
and $qq\bar q\bar Q$ tetraquark states in this paper. If the existence of the $X(5568)$ is confirmed, many other charmed/bottom tetraquarks may
also exist \cite{2016-Liu-p74023-74023,2016-Wang-p93101-93101,2016-Ali-p34036-34036}.
Hence, in this paper we shall investigate tetraquark systems with various spin-parity quantum numbers $J^P=0^+, 1^+, 2^+$ and $0^-, 1^-$
in the framework of QCD sum rules.

This paper is organized as follows. In Sect.~\Rmnum{2}, we construct the open-flavor charmed/bottom
tetraquark interpolating currents and introduce the QCD sum rule formalism. The two-point correlation
functions and the spectral densities are calculated for various channels. In Sect.~\Rmnum{3}, we establish
stable tetraquark mass sum rules and make reliable predictions for the mass spectra of these tetraquark
states. The last section is a brief summary.

\section{Formalism of tetraquark sum rules}\label{Sec:QSR}
In this section, we will briefly introduce the method of QCD sum
rules \cite{1979-Shifman-p385-447,1985-Reinders-p1-1,2000-Colangelo-p1495-1576} for the tetraquark
systems. To begin, we construct the diquark-antidiquark tetraquark operators with one heavy quark
and three light quark fields. In the diquark configurations, all models agree that only the color antisymmetric
scalar $q^T_a C\gamma_5q_b$ and axial-vector  $q^T_aC\gamma_\mu q_b$ diquark fields are favored to
maintain low color electrostatic field energy \cite{2005-Jaffe-p1-45}. To explore the lowest-lying tetraquarks,
we use only these two favored $S$-wave diquarks to compose the color antisymmetric
$[\mathbf{\bar 3_c}]_{q_1q_2} \otimes [\mathbf{3_c}]_{\bar{q_3}\bar Q}$ tetraquark currents following
Refs.~\cite{2013-Du-p14003-14003,2014-Chen-p54037-54037,2011-Chen-p34010-34010,2010-Chen-p105018-105018}
\begin{equation}
\begin{split}
J_1&=q^T_{1a}C\gamma_5q_{2b}(\bar{q}_{3a}\gamma_5C\bar{Q}^T_b-\bar{q}_{3b}\gamma_5C\bar{Q}^T_a)\, , ~~~~J^{P}=0^{+}\, ,\\
J_2&=q^T_{1a}C\gamma_\mu q_{2b}(\bar{q}_{3a}\gamma^\mu C\bar{Q}^T_b-\bar{q}_{3b}\gamma^\mu C\bar{Q}^T_a)\, , ~~~J^{P}=0^{+}\, ,\\
J_{3\mu}&=q^T_{1a}C\gamma_5q_{2b}(\bar{q}_{3a}\gamma_\mu C\bar{Q}_b^T-\bar{q}_{3b}\gamma_\mu C\bar{Q}_a^T)\, , ~~~~J^{P}=0^-, \, 1^{+}\, ,\\
J_{4\mu}&=q^T_{1a}C\gamma_\mu q_{2b}(\bar{q}_{3a}\gamma_5C\bar{Q}_b^T-\bar{q}_{3b}\gamma_5C\bar{Q}^T_a)\, , ~~~~J^{P}=0^-, \, 1^{+}\, ,\\
J_{5\mu\nu}&=q^T_{1a}C\gamma_\mu q_{2b}(\bar{q}_{3a}\gamma_\nu C\bar{Q}_b^T-\bar{q}_{3b}\gamma_\nu C\bar{Q}_a^T)\, ,
~~~~J^{P}=0^{+}, \, 1^-, \, 2^+ (\mbox{S}); \, 1^-, \, 1^+ (\mbox{A}); \, 0^+ (\mbox{T})\, ,
\label{currents}
\end{split}
\end{equation}
in which $Q$ is a heavy quark ($c$ or $b$) and $q_1, q_2, q_3$ are light quarks ($u, d, s$). For the tensor current $J_{5\mu\nu}$, we list its $J^P$ assignments for the traceless symmetric part (S), the antisymmetric part (A) and
the trace (T). All these interpolating currents can couple to the tetraquark states that carry the same spin-parity
quantum numbers. In this paper, we shall investigate the charm-strange $[sq][\bar q\bar c]$, non-strange
charmed $[qq][\bar q\bar c]$, bottom-strange $[sq][\bar q\bar b]$, and non-strange bottom $[qq][\bar q\bar b]$ tetraquark
systems by using the interpolating currents in Eq.~\eqref{currents}, where $q$ is an up or down quark.

We shall study the following two-point correlation functions using the scalar, vector and
tensor currents
\begin{align}
\Pi(q^2)&=i\int d^4x \,e^{iq\cdot x}\,\langle 0|T [J(x)J^\dagger(0)]|0\rangle\, , \label{piscalar}\\
\Pi_{\mu\nu}(q^2)&=i\int d^4x \,e^{iq\cdot x}\,\langle 0|T [J_\mu(x)J_\nu^\dagger(0)]|0\rangle\, , \label{pivector}\\
\Pi_{\mu\nu,\rho\sigma}(q^2)&=i\int d^4x\,e^{iq\cdot x}\,\langle0|T[J_{\mu\nu}(x)J_{\rho\sigma}^{\dag}(0)]|0\rangle\, . \label{pitensor}
\end{align}
In general, the two-point functions $\Pi_{\mu\nu}(q^2)$ in Eq.~\eqref{pivector} and $\Pi_{\mu\nu,\rho\sigma}(q^2)$
in Eq.~\eqref{pitensor} contain several different invariant functions referring to pure spin-0, spin-1 or spin-2 hadron
states. These invariant functions have distinct tensor structures in $\Pi_{\mu\nu}(q^2)$ and $\Pi_{\mu\nu,\rho\sigma}(q^2)$.
For the vector current, it is easy to write the corresponding two-point function as
 \begin{align}
\Pi_{\mu\nu}(q^2)=\eta_{\mu\nu}\Pi_V(q^2)+\frac{q_\mu q_\nu}{q^2}\Pi_S(q^2)\, , \label{pivector2}
\end{align}
where $\eta_{\mu\nu}=q_\mu q_\nu/q^2-g_{\mu\nu}$ is a projector for the pure spin-1 invariant function $\Pi_V(q^2)$
while $q_\mu q_\nu/q^2$ is a projector for spin-0 invariant function $\Pi_S(q^2)$. To pick out different invariant functions
in $\Pi_{\mu\nu,\rho\sigma}(q^2)$ for the tensor current $J_{5\mu\nu}$, we introduce some projectors following
Ref.~\cite{1987-Govaerts-p674-674}
\begin{equation}
\begin{split}
P_{0T}&=\frac{1}{16}g_{\mu\nu}g_{\rho\sigma}\, , ~~~~~~~\mbox{for}\, J^P=0^+,\, \mbox{T}\\
P_{0S}&=T_{\mu\nu}T_{\rho\sigma}\, , ~~~~~~~~~~\mbox{for}\, J^P=0^+,\, \mbox{S}\\
P_{0TS}&=\frac{1}{4}(T_{\mu\nu}g_{\rho\sigma}+T_{\rho\sigma}g_{\mu\nu})\, , ~~~~~~~~~~\mbox{for}\, J^P=0^+,\, \mbox{TS}\\
P_{1A}^N&=T_{\mu\nu,\rho\sigma}^-\, , ~~~~~~~\mbox{for}\, J^P=1^-,\, \mbox{A}\\
P_{1S}^N&=T_{\mu\nu,\rho\sigma}^+\, , ~~~~~~~\mbox{for}\, J^P=1^-,\, \mbox{S}\\
P_{1AS}^N&=2\left(\frac{q_\mu q_\rho}{q^2}\eta_{\nu\sigma}-\frac{q_\nu q_\sigma}{q^2}\eta_{\mu\rho}\right)\, , ~~~~~~~\mbox{for}\, J^P=1^-,\, \mbox{AS}\\
P_{1A}^P&=\eta_{\mu\rho}\eta_{\nu\sigma}-\eta_{\mu\sigma}\eta_{\nu\rho}\, , ~~~~~~~\mbox{for}\, J^P=1^+,\, \mbox{A}\\
P_{2S}^N&=\left(\eta_{\mu\rho}\eta_{\nu\sigma}+\eta_{\mu\sigma}\eta_{\nu\rho}-\frac{2}{3}\eta_{\mu\nu}\eta_{\rho\sigma}\right)\, , ~~~~~~~\mbox{for}\, J^P=2^+,\, \mbox{S} \label{projectors}
\end{split}
\end{equation}
where
\begin{equation}
\begin{split}
T_{\mu\nu}&=\frac{q_\mu q_\nu}{q^2}-\frac{1}{4}g_{\mu\nu}\, , \\
T_{\mu\nu,\rho\sigma}^\pm&=\left[\frac{q_\mu q_\rho}{q^2}\eta_{\nu\sigma}\pm(\mu\leftrightarrow\nu)\right]\pm(\rho\leftrightarrow\sigma)\, .
\end{split}
\end{equation}
One notes that in Eq.~\eqref{projectors} there are three different projectors for the scalar ($J^P=0^+$) channel, $P_{0T}$, $P_{0S}$
and $P_{0TS}$, which can be used to pick out different invariant functions induced by the the trace part,
traceless symmetric part and the cross term respectively from the current $J_{5\mu\nu}$. However, all these
three invariant functions couple to the $J^P=0^+$ channel with different coupling constants. We will
discuss all of them in this paper. A similar situation happens for the vector ($J^P=1^-$) channel in
Eq.~\eqref{projectors}.

As usual, the dispersion relation is used to describe the two-point correlation function at the hadronic level
\begin{eqnarray}
\Pi(q^2)=\frac{(q^2)^N}{\pi}\int_{s<}^{\infty}\frac{\mbox{Im}\Pi(s)}{s^N(s-q^2-i\epsilon)}ds+\sum_{n=0}^{N-1}b_n(q^2)^n\, ,
\label{Phenpi}
\end{eqnarray}
in which the unknown subtraction constants $b_n$ in the second term can be removed by taking the Borel
transform of $\Pi(q^2)$. The imaginary part of the two-point function is usually defined as the spectral function,
which can be evaluated at the hadronic level by inserting intermediate hadron states
\begin{align}
\rho(s)\equiv\frac{\mbox{Im}\Pi(s)}{\pi}&=\sum_n\delta(s-m_n^2)\langle0|J_{\mu}|n\rangle\langle n|J_{\mu}^{\dagger}|0\rangle \\
&=f_X^2\delta(s-m_X^2)+ \mbox{continuum},  \label{Phenrho}
\end{align}
where the single narrow pole plus continuum parametrization is adopted in the last step. The inserted intermediate
states $n$ carry the same quantum numbers as the interpolating current $J_{\mu}(x)$. The quantity $m_X$ denotes
the mass of the lowest lying resonance and $f_X$ is the coupling constant.
For the scalar and vector currents, the leptonic coupling constants are defined as
\begin{align}
\langle0|J|X\rangle&=f_S\,, \label{scalarcoupling}\\
\langle0|J_{\mu}|X\rangle&=f_V \epsilon_{\mu}+f_S\, q_\mu\,, \label{vectorcoupling}
\end{align}
in which $\epsilon_\mu$ is the polarization vector.
For the tensor current, the coupling constants are defined as
\begin{align}
\begin{split}
\langle0|J_{\mu\nu}|X\rangle&=f_{0T}g_{\mu\nu}+f_{0S}q_\mu q_\nu \, ~~~(J^P=0^+)\\
&+f_{1S}^-(\epsilon_\mu q_\nu+\epsilon_\nu q_\mu)+f_{1A}^-(\epsilon_\mu q_\nu-\epsilon_\nu q_\mu)\, ~~~(J^P=1^-) \\
&+f_{1A}^+ \epsilon^{\mu\nu\rho\sigma} \epsilon_\rho q_\sigma \, ~~~(J^P=1^+) \\
&+f_{2S} \epsilon_{\mu\nu}\, ~~~(J^P=2^+)\, ,
\end{split}
\end{align}
where $\epsilon_{\mu\nu}$ is the polarization tensor and $\epsilon^{\mu\nu\rho\sigma}$ is the completely
antisymmetric tensor.

At the quark-gluonic level, the correlation function can be computed by using perturbative QCD augmented with
nonperturbative quark and gluon condensates. Comparing the two-point correlation functions at both the
hadronic and quark-gluonic levels, one can establish QCD sum rules for hadron parameters like hadron masses,
magnetic moments and coupling constants. The technique of Borel transformation is usually adopted to remove the
unknown constants in Eq.~\eqref{Phenpi} and suppress the continuum contributions to correlation functions, which
results in the Borel sum rules
\begin{eqnarray}
\mathcal{L}_{k}(s_0, M_B^2)=f_X^2m_X^{2k}e^{-m_X^2/M_B^2}=\int_{M_Q^2}^{s_0}dse^{-s/M_B^2}\rho(s)s^k\, ,
\label{sumrule}
\end{eqnarray}
in which the lower integral limit $M_Q^2=(m_Q+m_{q_1}+m_{q_2}+m_{q_3})^2$ denotes a physical threshold for the
$q_1q_2\bar q_3\bar Q$ system. The mass of the lowest-lying hadron state is thus obtained as
\begin{eqnarray}
m_X(s_0, M_B^2)=\sqrt{\frac{\mathcal{L}_{1}(s_0, M_B^2)}{\mathcal{L}_{0}(s_0, M_B^2)}}\, , \label{mass}
\end{eqnarray}
where $M_B$ is the Borel parameter and $s_0$ is the continuum threshold above which the contributions from
the continuum and higher excited states can be approximated well by the QCD spectral function $\rho(s)$.

In this paper, we calculate the correlation functions and spectral densities by considering the perturbative term, quark condensates, gluon condensate, quark-gluon mixed condensates, four-quark condensates and the dimension eight condensates  at leading order in $\alpha_s$. In our evaluation, the strange quark propagator was considered in
momentum space. For all interpolating currents in Eq.~\eqref{currents}, we collect the expressions of the spectral densities in Appendix \ref{sec:rhos}.  We use the projectors defined in Eq.~\eqref{projectors} to pick out the different invariant functions for the vector and tensor currents. As shown in Eq.~\eqref{currents}, the tensor current $J_{5\mu\nu}$
can couple to two different scalar channels ($J^P=0^+$) as well as two different vector channels ($J^P=1^-$).
We evaluate the invariant functions and spectral densities for all these channels and list them in Appendix
\ref{sec:rhos}. In addition, one can also find the projectors $P_{0TS}$ (for $J^P=0^+$ TS) and $P_{1AS}^N$
(for $J^P=1^-$ AS) in Eq.~\eqref{projectors}, which can be used to pick out the cross-term invariant functions
for the scalar and vector channels respectively. However, we find that the perturbative terms for these two
invariant functions are proportional to the light quark mass $m_q$, which can be neglected in the chiral limit.
Such invariant functions can not provide reliable predictions for hadron properties and thus we will not use
them to perform QCD sum rule analyses in this paper.

\section{QCD sum rule analysis}\label{Sec:NA}
In this section, we use the two-point correlation functions obtained above to perform QCD sum rule analyses for the
charmed/bottom $su\bar d\bar Q$ and $ud\bar d\bar Q$ tetraquark systems, in which the following parameter values
for the quark masses and QCD condensates are adopted
\cite{1985-Reinders-p1-1,2016-Patrignani-p100001-100001,2012-Narison-p259-263,2007-Kuhn-p192-215}
\begin{equation}
\begin{split}
& m_u=m_d=0\, , \\
& m_s(2\,\text{GeV})=(96^{+8}_{-4})\text{ MeV} \, ,
\\ &
m_c(m_c)=\overline m_c=(1.27\pm 0.03) \mbox{ GeV}   \, ,
\\ &
m_b(m_b)=\overline m_b=(4.18^{+0.04}_{-0.03}) \mbox{ GeV}   \, ,
\\&
\qq=-(0.24\pm0.01)^3\text{ GeV}^3 \, ,
\\&
\ss=(0.8\pm0.1)\qq \, ,
\\ &
\qGqa=-M_0^2\qq \, ,
\\ &
\sGsa=-M_0^2\ss \, ,
\\ &
M_0^2=(0.8\pm0.2)\text{ GeV}^2 \, ,
\\ &
\GGb=(0.48\pm0.14) \text{ GeV}^4\, , \label{parameters}
\end{split}
\end{equation}
where we use the ``running masses" for the heavy quarks in the $\overline{\rm MS}$ scheme.
We use the chiral limit in our analysis in which the up and down quark masses $m_u=m_d=m_q=0$.

As shown in Eq.~\eqref{mass}, the extracted hadron mass is a function of the Borel parameter
$M_B$ and the continuum threshold $s_0$. To obtain a reliable mass sum rule analysis, one should
choose suitable working ranges for these two free parameters. We determine the lower bound on
the Borel parameter by requiring the perturbative term contribution be larger than three times of
the dominant nonperturbative terms. The study of the pole contribution can yield the upper bound
on $M_B$. For the continuum threshold $s_0$, we will choose a reasonable value to minimize
the dependance of the extracted hadron mass with respect to $M_B$. In the following, we will use
the $sq\bar q\bar c$ systems as examples to discuss the detail of the numerical analyses for all
channels, which can be classified into three types:

\begin{enumerate}

\item[A.] This type can provide stable mass sum rules and give reliable mass predictions. We use the
interpolating current $J_2(x)$ with $J^{P}=0^{+}$ as an example to perform the numerical analysis.
As shown in the left panel of Fig.~\ref{fig:masscsud0+2}, we show the variation of $m_X$ with respect
to the continuum threshold $s_0$ for different value of the Borel mass $M_B^2$. We find that there are
some minimum points around which $m_X$ is stable at $s_0=7.0$ GeV$^2$. For larger continuum
threshold after these points, the hadron mass will increase gradually with $s_0$ and the curves with
different values of $M_B^2$ intersect at $s_0=10.0$ GeV$^2$. Thus we can determine the optimal
working range for the continuum threshold with a reasonable $10\%$ uncertainty to be $9.0\leq s_0\leq
11.0$ GeV$^2$ (shaded region in the left panel of Fig.~\ref{fig:masscsud0+2}), in which the $M_B$
dependance of $m_X$ will be very weak. Accordingly, we can also obtain the Borel window as
$4.7\leq M_B^2\leq 5.0$ GeV$^2$ by studying the OPE convergence and pole contribution. We show
the Borel curves in the right panel of Fig.~\ref{fig:masscsud0+2} for the hadron mass $m_X$. One notes
that the sum rules are very stable in the Borel window and the extracted hadron mass increases with
respect to $s_0$. Using the central value $s_0=10.0$ GeV$^2$, we can extract the hadron mass for
the $J^{P}=0^{+}$ $sq\bar q\bar c$ tetraquark state
\begin{equation}
m_{X_{0^+,\, 2}^{cs}}=(2.91\pm0.11\pm0.07\pm0.04\pm0.01)\, \mbox{GeV}\, , \label{mass0+2}
\end{equation}
in which the errors come from the uncertainties in the threshold values $s_0$, $M_0^2$, various QCD
condensates and the charm quark mass, respectively. 
\begin{figure*}[hbt]
\begin{center}
\scalebox{0.63}{\includegraphics{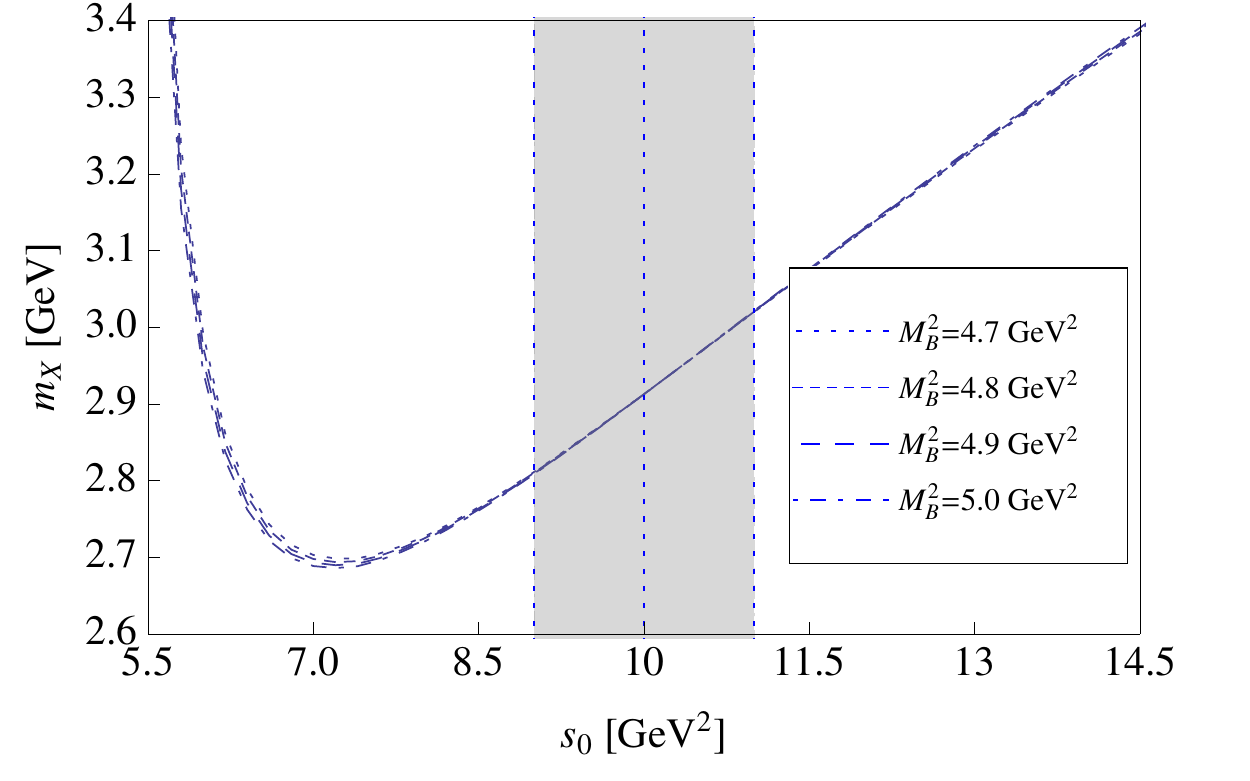}}
\scalebox{0.63}{\includegraphics{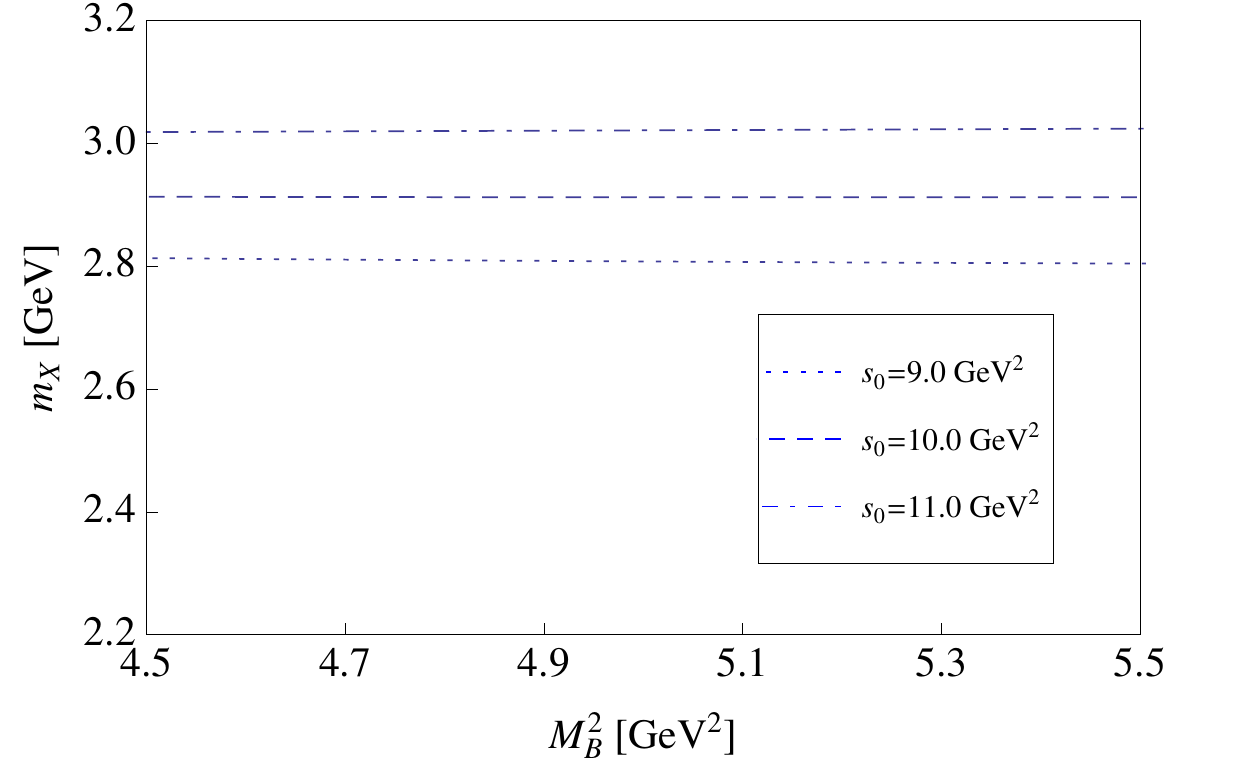}}
\caption{Variation of the hadron mass $m_X$ with respect to $s_0$ and $M_B^2$ for $sq\bar q\bar c$ system with
$J^{P}=0^{+}$ using the interpolating current $J_2(x)$.
}
\label{fig:masscsud0+2}
\end{center}
\end{figure*}
\begin{figure*}[hbt]
\begin{center}
\scalebox{0.63}{\includegraphics{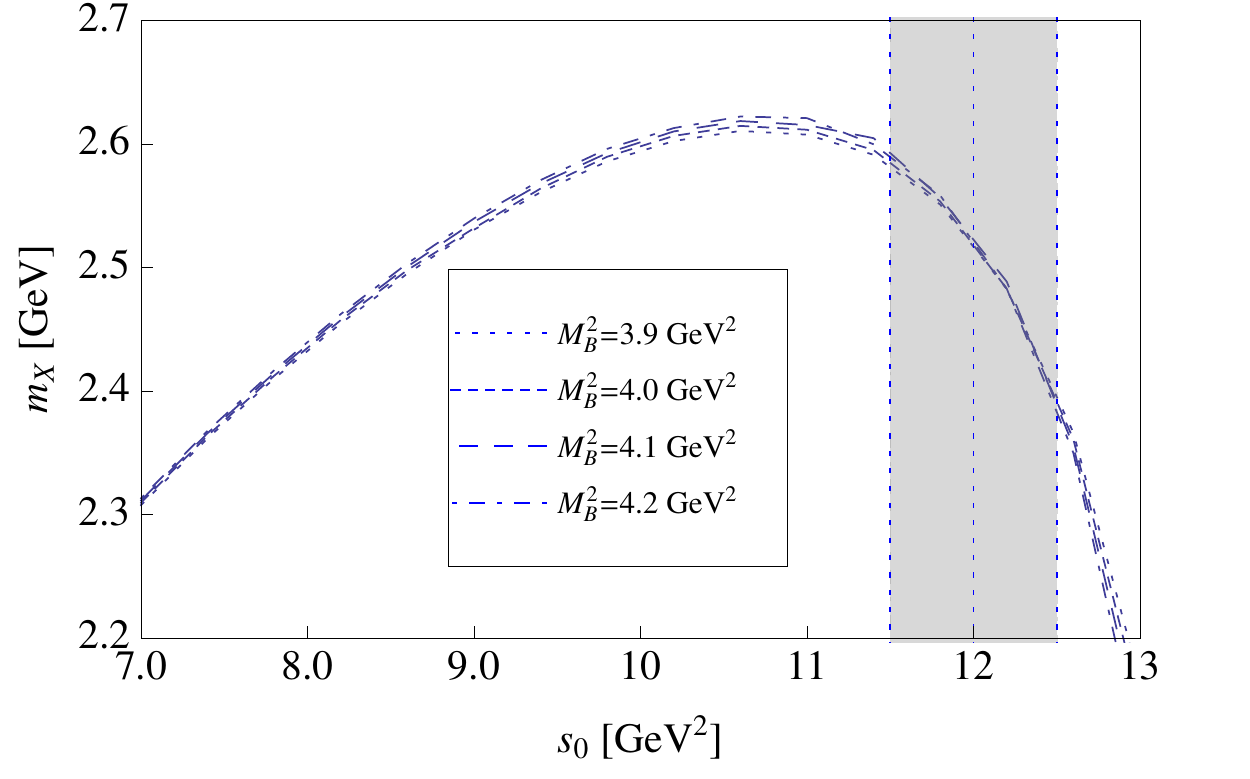}}
\scalebox{0.63}{\includegraphics{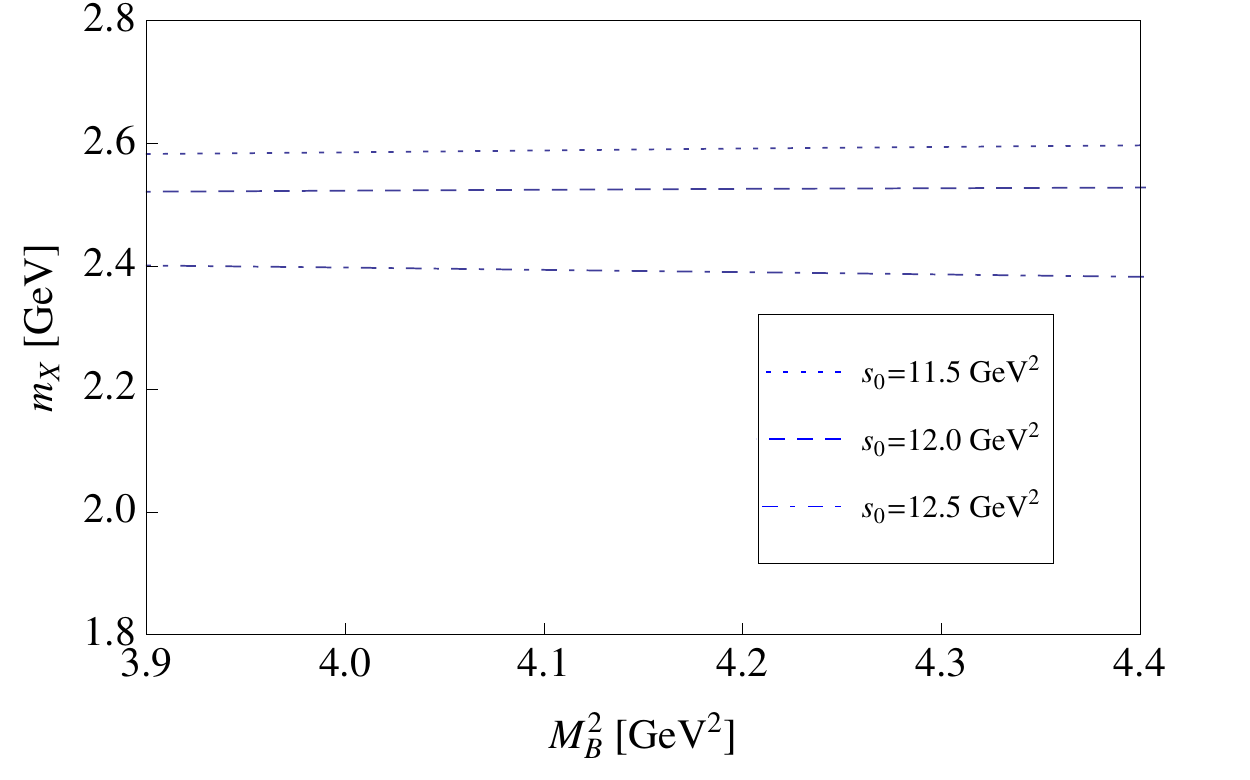}}
\caption{Variation of the hadron mass $m_X$ with respect to $s_0$ and $M_B^2$ for $sq\bar q\bar c$ system
with $J^{P}=0^{+}$ (S) using the interpolating current $J_{5\mu\nu}(x)$.}
\label{fig:masscsud0+S}
\end{center}
\end{figure*}

\item [B.] In this type we consider the traceless symmetric part of the interpolating current $J_{5\mu\nu}(x)$
in the scalar channel with $J^{P}=0^{+}$ (S). In the left panel of Fig . \ref{fig:masscsud0+S}, we show the
variation of the $m_X$ with respect to the continuum threshold $s_0$ in the Borel window $3.9\leq M_B^2\leq
4.2$ GeV$^2$. We find that the behaviour of these $s_0$-dependance curves is very different from those in
type A as shown in Fig . \ref{fig:masscsud0+2}. Instead of minimum points for type A, the $s_0$-dependance
curves in type B have maximum points and then the extracted hadron mass decreases
gradually with respect to $s_0$. However, we are still able to find the optimal values for the continuum threshold
$11.5\leq s_0\leq 12.5$ GeV$^2$ to minimize the dependance of $m_X$ on the Borel mass $M_B^2$. In this
working range, we plot the stable Borel curves in the right panel of Fig.~\ref{fig:masscsud0+S} in the above
Borel window, from which the extracted hadron mass decreases with respect to $s_0$. We finally obtain
\begin{equation}
m_{X_{0^+,\, S}^{cs}}=(2.52\pm0.10\pm0.11\pm0.06\pm0.05)\, \mbox{GeV}\, , \label{mass0+S}
\end{equation}
in which the errors come from the uncertainties in the threshold values $s_0$, $M_0^2$, various QCD
condensates and the charm quark mass, respectively. 

\item [C.] In the third type we study the traceless antisymmetric part of the interpolating current $J_{5\mu\nu}(x)$
in the vector channel with $J^{P}=1^{-}$ (A). We first study the variation of the hadron mass with $s_0$ in its Borel
window $2.6\leq M_B^2\leq 3.2$ GeV$^2$. As shown in the left panel of Fig.~\ref{fig:masscsud1-A}, the behaviour
is totally different from those in types A and B. The extracted hadron mass $m_X$ increases monotonically with $s_0$
without any minimum or maximum point and the curves with different value of $M_B^2$ do not intersect anywhere.
It seems that the mass sum rules are unstable in this situation. To explore the further behavior of $s_0$-dependance,
we define the following hadron mass
\begin{equation}
\overline m_X(s_0)=\sum_{i=1}^N\frac{m_X(s_0, M_{B,\ i}^2)}{N}\, , \label{averagemass}
\end{equation}
in which the $M_{B,\ i}^2 (i=1, 2, \cdots, N)$ represent $N$ definite values for the Borel parameter $M_B^2$ in the
Borel window $2.6\leq M_B^2\leq 3.2$ GeV$^2$. The $\overline m_X(s_0)$ is defined as an averaged hadron mass
for some definite value $M_B^2$. Using this average hadron mass, we can define the following quantity
\begin{equation}
\chi^2(s_0)=\sum_{i=1}^N\left[\frac{m_X(s_0, M_{B,\ i}^2)}{\overline m_X(s_0)}-1\right]^2\, . \label{chi}
\end{equation}
According to the above definition, the optimal choice for the continuum threshold $s_0$ in the QCD sum rule analysis
can be obtained by minimizing the quantity $\chi^2(s_0)$, which is only the function of $s_0$. We show this relation in
the right panel of Fig.~\ref{fig:masscsud1-A}, from which there is a minimum point around $s_0=13.5$ GeV$^2$. It is
clearly that the $M_B^2$-dependance for the extracted hadron mass is the weakest at this point. We can thus determine
the working range for the continuum threshold to be $12.0\leq s_0\leq 15.0$ GeV$^2$ in our analysis, as shown in the
left panel of Fig.~\ref{fig:masscsud1-A}. In this area, we show $m_X$ as a function of the Borel parameter $M_B^2$
in Fig.~\ref{fig:masscsud1-AMB} and predict the hadron mass at the central values $s_0=13.5$ GeV$^2$, $M_B^2=2.9$ GeV$^2$ to be
\begin{equation}
m_{X_{1^-,\, A}^{cs}}=(3.35\pm0.14\pm0.07\pm0.04\pm0.01)\, \mbox{GeV}\, , \label{mass1-A}
\end{equation}
in which the errors come from the uncertainties in the threshold values $s_0$, $M_0^2$, various QCD
condensates and the charm quark mass, respectively. 
\begin{figure*}[hbt]
\begin{center}
\scalebox{0.63}{\includegraphics{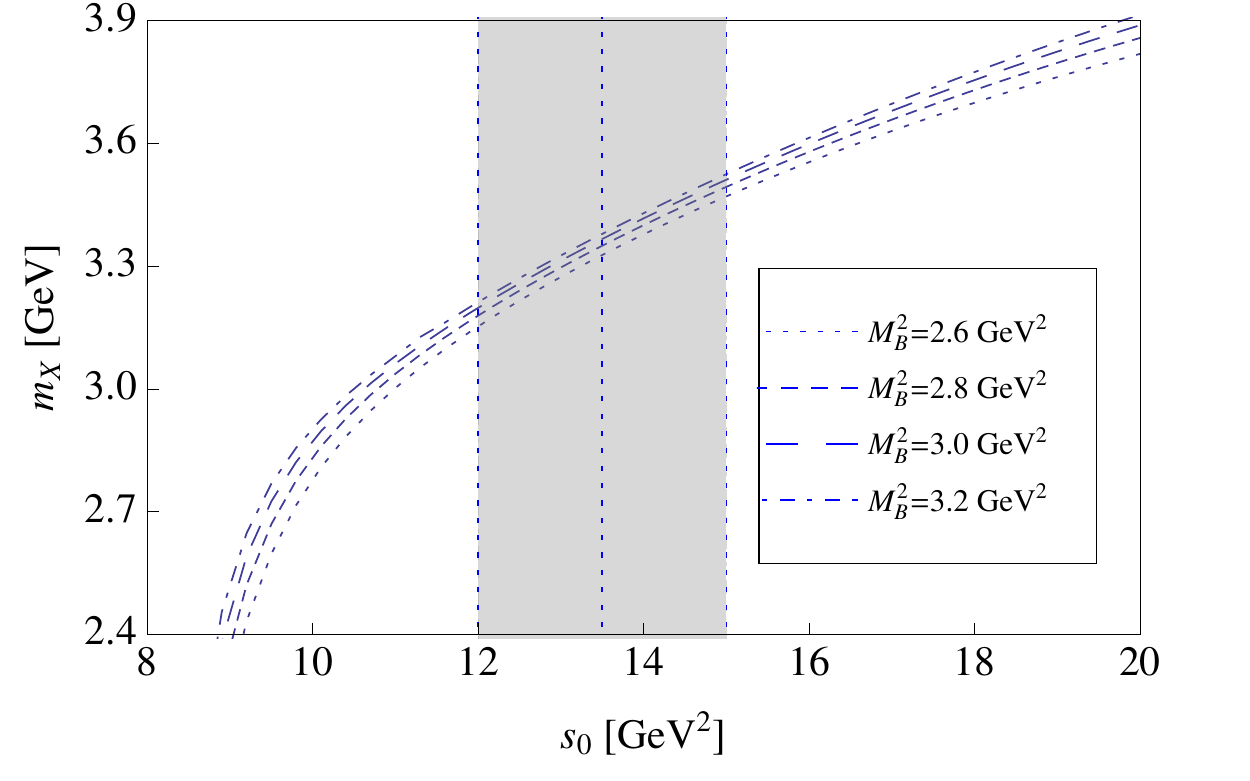}}
\scalebox{0.63}{\includegraphics{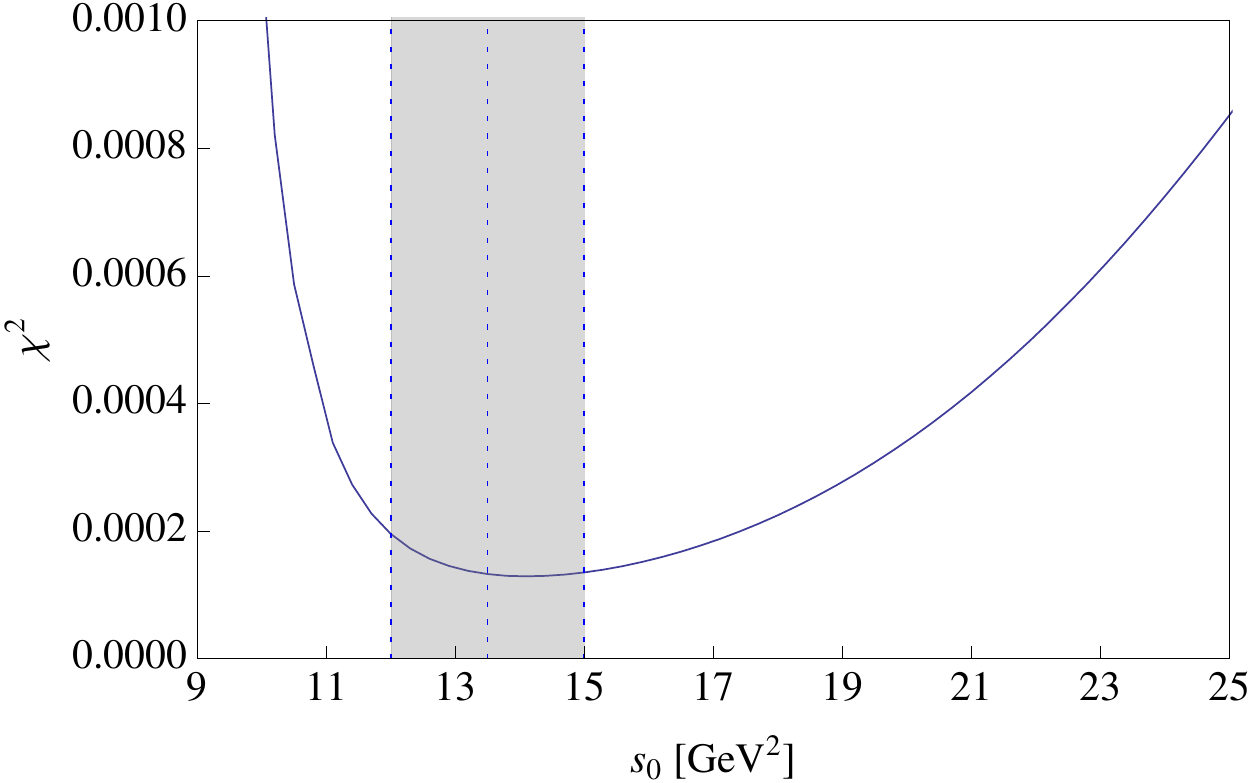}}
\caption{Left panel: hadron mass $m_X$ as a function of $s_0$ for the $sq\bar q\bar c$ system using the
traceless antisymmetric part of $J_{5\mu\nu}(x)$ with $J^{P}=1^{-}$(A). Right panel: the quantity
$\chi^2$ as the function of the continuum threshold $s_0$.}
\label{fig:masscsud1-A}
\end{center}
\end{figure*}
\begin{figure*}[hbt]
\begin{center}
\scalebox{0.65}{\includegraphics{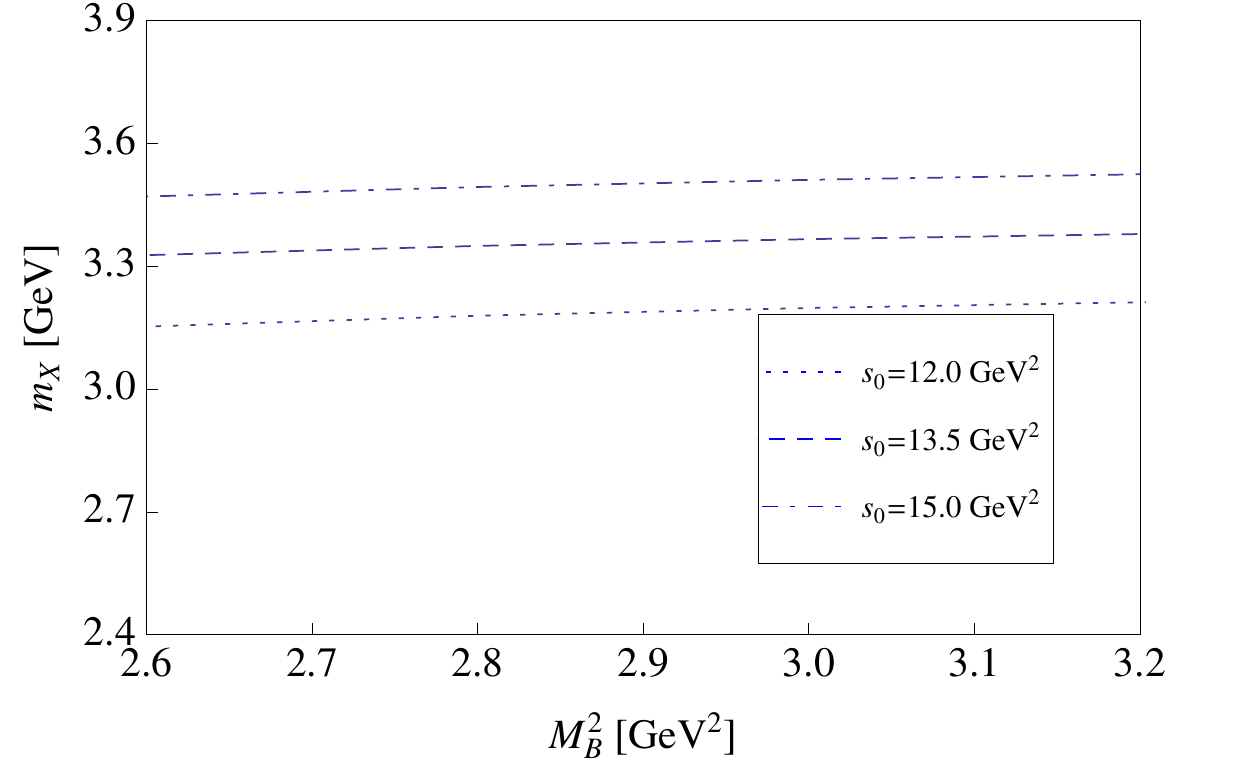}}
\caption{Mass prediction for the $sq\bar q\bar c$ system with $J^{P}=1^{-}$(A).}
\label{fig:masscsud1-AMB}
\end{center}
\end{figure*}
\end{enumerate}

\begin{table}
\renewcommand{\arraystretch}{1.3}
\begin{tabular*}{11.3cm}{clcccc}
\hlinewd{.8pt}
$J^{PC}$    ~  & Currents          ~ & ~ $s_0$\mbox{(GeV$^2$)}  ~ &~ \mbox{Borel window (GeV$^2$)}~
& ~$m_{X^{cs}}$\mbox{(GeV)} ~&~~~Type~~~  \\
\hline
$0^{+}$   & $J_1$                       &  $7.5\pm0.5$            & $3.3-3.6$         & $2.55\pm0.10$    & A\\
               & $J_2$                       &  $10.0\pm0.5$          & $4.7-5.0$         & $2.91\pm0.14$    & A\\
               & $J_{5\mu\nu}$(T)     &  $10.0\pm0.5$          & $2.2-3.1$         & $2.88\pm0.15$    & A\\
               & $J_{5\mu\nu}$(S)     &  $12.0\pm0.5$          & $3.9-4.2$         & $2.53\pm0.13$    & B
\vspace{.2cm}\\
$1^{+}$   & $J_{3\mu}$               &  $7.5\pm0.5$            & $3.5-3.8$         & $2.55\pm0.12$    & A\\
               & $J_{4\mu}$               &  $9.5\pm0.5$            & $3.2-3.9$         & $2.81\pm0.13$    & A\\
               & $J_{5\mu\nu}$(A)     &  $9.5\pm0.5$            & $3.5-3.8$         & $2.83\pm0.13$    & A
\vspace{.2cm}\\
$2^{+}$   & $J_{5\mu\nu}$(S)     &  $10.0\pm0.5$          & $3.4-4.0$         & $2.91\pm0.13$    & A
\vspace{.2cm}\\
$0^{-}$    & $J_{3\mu}$               &  $7.5\pm0.5$            & $4.0-4.3$         & $2.31\pm0.09$    & A\\
               & $J_{4\mu}$               &  $12.0\pm0.5$          & $3.2-4.1$         & $3.30\pm0.16$    & A
\vspace{.2cm}\\
$1^{-}$    & $J_{5\mu\nu}$(A)     &  $13.5\pm1.0$          & $2.6-3.4$         & $3.35\pm0.16$    & C\\
               & $J_{5\mu\nu}$(S)     &  $15.5\pm1.0$          & $3.2-3.9$         & $3.55\pm0.18$    & C\\
\hlinewd{.8pt}
\end{tabular*}
\caption{Mass spectrum for the charm-strange $sq\bar q\bar c$ tetraquark states. \label{csudmass}}
\end{table}

For all interpolating currents in Eq.~\eqref{currents}, we perform similar numerical analyses and collect the
extracted hadron masses for the $sq\bar q\bar c$ tetraquark states in Table \ref{csudmass}, together with
the Borel windows and the working ranges for $s_0$. We show the three types introduced above in the
last column. The error sources for the hadron masses include the uncertainties of the heavy quark masses,
the QCD condensates, $M_0^2$, and the uncertainty of the continuum threshold $s_0$. As shown
in Eqs.~\eqref{mass0+2}, \eqref{mass0+S} and \eqref{mass1-A}, the uncertainty in $s_0$ is the dominant error source
of the hadron mass while that of $M_0^2$ parameterizing the mixed condensate is also important. However, we list only the total errors in Table \ref{csudmass} with error analyses to summarize the results. 
In Table \ref{csudmass}, we find that the extracted masses for the scalar $sq\bar q\bar c$ charmed tetraquarks
with $J^P=0^+$ to be almost degenerate around $2.54$ GeV from the currents $J_1(x)$ and $J_{5\mu\nu}(x)$(S),
while $2.90$ GeV from the currents $J_2(x)$ and $J_{5\mu\nu}(x)$(T). These values for the scalar charmed
tetraquarks are higher than the mass of the charm-strange $D_{s0}^\ast(2317)$ meson. In addition,
we obtain the hadron mass $m_{X_{cs}}=(2.55\pm0.12)$ GeV for the axial-vector $sq\bar q\bar c$
tetraquark using the current $J_{3\mu}(x)$ with $J^P=1^+$. This result is not far from the mass of the
narrow charm-strange $D_{s1}(2460)$ meson within the error.

Replacing the strange quark in $sq\bar q\bar c$ systems to be a down quark, we can study the non-strange
charmed tetraquark systems $qq\bar q\bar c$ in similar way as the above analyses. The OPE series are a
bit different by changing the condensates $\ss$, $\sGsb$ into $\qq$, $\qGqb$ respectively and neglecting
the $m_q$ proportional terms in the chiral limit. The numerical results for these systems are then obtained
and collected in Table \ref{cqqqmass}. Similarly, we can easily study the strange and non-strange bottom
tetraquark systems $sq\bar q\bar b$ and $qq\bar q\bar b$ in the heavy quark symmetry. After performing
the QCD sum rule analyses, we collect the numerical results for the $sq\bar q\bar b$ and $qq\bar q\bar b$
systems in Tables \ref{bsudmass} and \ref{bqqqmass}, respectively.
In Table \ref{bsudmass}, the masses for the bottom-strange $sq\bar q\bar b$ tetraquarks extracted from the
interpolating currents $J_1(x)$ and $J_{3\mu}(x)$ were previously obtained in Ref.~\cite{2016-Chen-p22002-22002},
which were used to explain the newly reported $X(5568)$ structure.

Except for the scalar and axial-vector states, we also investigate the other channels with $J^P=0^-, 1^-, 2^+$ and
collect the results in Tables \ref{csudmass}-\ref{bqqqmass}.

\begin{table}
\renewcommand{\arraystretch}{1.3}
\begin{tabular*}{11.3cm}{clcccc}
\hlinewd{.8pt}
$J^{PC}$    ~  & Currents          ~ & ~ $s_0$\mbox{(GeV$^2$)}  ~ &~ \mbox{Borel window (GeV$^2$)}~
& ~$m_{X^{cq}}$\mbox{(GeV)} ~&~~~Type~~~  \\
\hline
$0^{+}$   & $J_1$                &  $9.5\pm0.5$              & $3.1-4.0$         & $2.82\pm0.12$    & A\\
               & $J_2$                &  $10.0\pm0.5$            & $4.4-5.0$         & $2.91\pm0.13$    & A\\
               & $J_{5\mu\nu}$(T)     &  $10.5\pm0.5$     & $2.1-2.6$         & $2.93\pm0.13$    & A\\
               & $J_{5\mu\nu}$(S)     &  $8.5\pm0.5$       & $2.4-3.3$         & $2.23\pm0.12$    & B
\vspace{.2cm}\\
$1^{+}$        & $J_{3\mu}$           &  $9.5\pm0.5$      & $3.3-3.9$         & $2.83\pm0.12$    & A\\
               & $J_{4\mu}$                &  $10.0\pm0.5$    & $3.0-4.2$         & $2.88\pm0.12$    & A\\
               & $J_{5\mu\nu}$(A)      &  $9.5\pm0.5$      & $3.3-3.9$         & $2.82\pm0.12$    & A
\vspace{.2cm}\\
$2^{+}$       & $J_{5\mu\nu}$(S) &  $10.0\pm0.5$     & $3.2-4.1$         & $2.90\pm0.15$    & A
\vspace{.2cm}\\
$0^{-}$        & $J_{3\mu}$           &  $6.5\pm0.5$       & $4.1-4.4$         & $2.21\pm0.06$    & A\\
               & $J_{4\mu}$               &  $13.0\pm0.5$     & $3.4-4.0$         & $3.29\pm0.15$    & C
\vspace{.2cm}\\
$1^{-}$        & $J_{5\mu\nu}$(A)  &  $13.5\pm1.0$    & $2.7-3.3$         & $3.29\pm0.16$    & C\\
               & $J_{5\mu\nu}$(S)      &  $15.5\pm1.0$    & $3.3-4.0$         & $3.52\pm0.17$    & C\\
\hlinewd{.8pt}
\end{tabular*}
\caption{Mass spectrum for the non-strange charmed $qq\bar q\bar c$ tetraquark states. \label{cqqqmass}}
\end{table}
\begin{table}
\renewcommand{\arraystretch}{1.3}
\begin{tabular*}{11.3cm}{clcccc}
\hlinewd{.8pt}
$J^{PC}$    ~  & Currents          ~ & ~ $s_0$\mbox{(GeV$^2$)}  ~ &~ \mbox{Borel window (GeV$^2$)}~
& ~$m_{X^{bs}}$\mbox{(GeV)} ~&~~~Type~~~  \\
\hline
$0^{+}$   & $J_1$                       &  $34.0\pm2.0$            & $6.0-6.3$         & $5.59\pm0.18$    & A\\
               & $J_2$                       &  $37.0\pm2.0$            & $8.4-8.7$         & $5.83\pm0.21$    & A\\
               & $J_{5\mu\nu}$(T)     &  $41.0\pm2.0$            & $4.3-5.3$         & $6.02\pm0.21$    & A\\
               & $J_{5\mu\nu}$(S)     &  $61.0\pm2.0$            & $6.1-7.3$         & $7.32\pm0.21$    & A
\vspace{.2cm}\\
$1^{+}$        & $J_{3\mu}$          &  $34.0\pm2.0$            & $6.3-6.6$         & $5.59\pm0.19$    & A\\
               & $J_{4\mu}$               &  $38.0\pm2.0$            & $6.0-7.3$         & $5.86\pm0.20$    & A\\
               & $J_{5\mu\nu}$(A)     &  $38.0\pm2.0$            & $6.3-7.4$         & $5.88\pm0.18$    & A
\vspace{.2cm}\\
$2^{+}$        & $J_{5\mu\nu}$(S)&  $40.0\pm2.0$            & $6.2-8.0$         & $6.01\pm0.22$    & A
\vspace{.2cm}\\
$0^{-}$        & $J_{3\mu}$           &  $38.0\pm2.0$            & $6.6-6.9$         & $5.62\pm0.24$    & A\\
                   & $J_{4\mu}$           &  $43.0\pm2.0$            & $5.5-6.7$         & $6.32\pm0.18$    & A
\vspace{.2cm}\\
$1^{-}$        & $J_{5\mu\nu}$(A) &  $43.0\pm2.0$            & $4.4-5.0$         & $6.19\pm0.25$    & C\\
               & $J_{5\mu\nu}$(S)     &  $44.0\pm2.0$            & $4.8-5.5$         & $6.22\pm0.27$    & C\\
\hlinewd{.8pt}
\end{tabular*}
\caption{Mass spectrum for the bottom-strange $sq\bar q\bar b$ tetraquark states. \label{bsudmass}}
\end{table}
\begin{table}
\renewcommand{\arraystretch}{1.3}
\begin{tabular*}{11.3cm}{clcccc}
\hlinewd{.8pt}
$J^{PC}$    ~  & Currents          ~ & ~ $s_0$\mbox{(GeV$^2$)}  ~ &~ \mbox{Borel window (GeV$^2$)}~
& ~$m_{X^{bq}}$\mbox{(GeV)} ~&~~~Type~~~  \\
\hline
$0^{+}$   & $J_1$                       &  $38.0\pm2.0$            & $5.8-7.4$         & $5.86\pm0.20$    & A\\
               & $J_2$                       &  $37.0\pm2.0$            & $8.0-8.3$         & $5.83\pm0.20$    & A\\
               & $J_{5\mu\nu}$(T)     &  $41.0\pm2.0$            & $4.3-5.3$         & $6.02\pm0.21$    & A\\
               & $J_{5\mu\nu}$(S)     &  $52.0\pm2.0$            & $5.1-5.6$         & $6.74\pm0.24$    & C
\vspace{.2cm}\\
$1^{+}$        & $J_{3\mu}$          &  $38.0\pm2.0$            & $6.1-7.4$         & $5.87\pm0.21$    & A\\
               & $J_{4\mu}$               &  $40.0\pm2.0$            & $5.8-6.7$         & $5.98\pm0.21$    & A\\
               & $J_{5\mu\nu}$(A)     &  $38.0\pm2.0$            & $6.1-7.4$         & $5.86\pm0.21$    & A
\vspace{.2cm}\\
$2^{+}$        & $J_{5\mu\nu}$(S)&  $40.0\pm2.0$            & $6.0-6.7$         & $5.99\pm0.21$    & A
\vspace{.2cm}\\
$0^{-}$        & $J_{3\mu}$           &  $37.0\pm2.0$            & $6.8-7.1$         & $5.51\pm0.25$    & A\\
               & $J_{4\mu}$               &  $44.0\pm2.0$            & $5.6-6.0$         & $6.23\pm0.24$    & C
\vspace{.2cm}\\
$1^{-}$        & $J_{5\mu\nu}$(A) &  $40.0\pm2.0$            & $4.1-4.7$         & $5.88\pm0.27$    & C\\
               & $J_{5\mu\nu}$(S)     &  $41.0\pm2.0$            & $4.5-5.0$         & $5.94\pm0.27$    & C\\
\hlinewd{.8pt}
\end{tabular*}
\caption{Mass spectrum for the non-strange bottom $qq\bar q\bar b$ tetraquark states. \label{bqqqmass}}
\end{table}

\section{Decay properties of the open-flavor charmed/bottom tetraquarks}\label{Sec:decays}

Using the mass spectra obtained above, we can study the possible decay patterns of the $sq\bar q\bar c$,
$qq\bar q\bar c$, $sq\bar q\bar b$, $qq\bar q\bar b$ tetraquark states in various channels. These open-flavor
charmed/bottom tetraquarks will decay easily through the fall-apart mechanism so long as the kinematics allows.
We study both the S-wave and P-wave two-body hadronic decays by considering the conservation of the angular
momentum, parity and isospin in Tables \ref{S-wavedecays} and \ref{P-wavedecays}.

In Table \ref{S-wavedecays}, we list the possible S-wave two-body hadronic decay modes for the $sq\bar q\bar c$,
$qq\bar q\bar c$, $sq\bar q\bar b$, $qq\bar q\bar b$ tetraquark states with various quantum numbers. We consider
isospin-0/1 for $sq\bar q\bar c$, $sq\bar q\bar b$ states and isospin-$\frac{1}{2}$/$\frac{3}{2}$ for $qq\bar q\bar c$,
$qq\bar q\bar b$ states, respectively. In the chiral limit, these tetraquarks in the same isospin multiplet are predicted
to be degenerate since we do not differentiate between the up and down quarks. For the charmed/bottom-strange $sq\bar q\bar c$ and $sq\bar q\bar b$ states, their decay patterns are very different for the isospin-scalar and isospin-vector channels except some one $D/B$ meson plus one $K$ meson decay modes. Such decays are allowed by the isospin symmetry
for both channels. However, the situation is different for the non-strange $qq\bar q\bar c$ and $qq\bar q\bar b$ tetraquarks.
In Table \ref{S-wavedecays}, one notes that all possible decay modes for the isospin-$\frac{3}{2}$ states are allowed
for the corresponding isospin-$\frac{1}{2}$ ones.

As shown in Table \ref{S-wavedecays}, there is no allowed S-wave decay modes for the tensor $sq\bar q\bar b$ states.
This is because the predicted hadron mass for these tetraquarks in Table \ref{bsudmass} is lower than any possible
S-wave two-body hadronic decay threshold. There also exist some other tetraquark states below the S-wave decay
thresholds, which are denoted by ``$-$" in Table \ref{S-wavedecays}. However, it is shown that the P-wave decays are allowed for these states, as shown in Table \ref{P-wavedecays}. This means that the P-wave decay modes are dominant for these tetraquark states and thus they are much narrower than other states. These tetraquark states will be prime
candidates for observation.

\begin{table}
\renewcommand{\arraystretch}{1.5}
\begin{tabular*}{17cm}{ccc|ccc}
\hlinewd{.8pt}
$I(J^{P})$  ~~   &~~ $sq\bar q\bar b$    ~&~ $sq\bar q\bar c$~ & ~$I(J^{P})$  ~~   & $qq\bar q\bar b$  ~ & ~$qq\bar q\bar c$~  \\
\hline
$0(0^{+})$   & $B_s^0\eta/\eta^{\prime}$, $B_s^{\ast}\omega/\phi$, $B^\ast K^\ast(892)$,
& $D_s\eta/\eta^\prime$, &$\frac{1}{2}(0^{+})$   &$B\eta/\eta^{\prime}$, $B^{\ast}\omega/\phi$,  & $D\pi/\eta/\eta^\prime$,  \\
& $B_{s1}(5830)^0h_1(1170)/ f_1(1285)$, & $D_s^\ast\omega$,&&$B\pi$, $B^{\ast}\rho$&$D^\ast\omega/\rho$\\
& $B_{1}(5721)^0K_1(1270)$, $BK$&$DK$&&&\\
$1(0^{+})$   & $B_s^0\pi$, $BK$, $B_s^{\ast}\rho$, $B^\ast K^\ast(892)$,  & $D_s\pi$,&$\frac{3}{2}(0^{+})$   &$B\pi$, $B^{\ast}\rho$
& $D\pi$,  \\
& $B_{s1}(5830)^0b_1(1235)/ a_1(1260)$,&$D_s^\ast\rho$, &&&$D^\ast\rho$\\
& $B_{1}(5721)^0K_1(1270)$ &$DK$&&&\\
$0(1^{+})$   & $-$    &$-$&$\frac{1}{2}(1^{+})$   &$B^{\ast}\pi$, $B\omega$ & $D^{\ast}\pi$, $D\omega/\rho$  \\
$1(1^{+})$   & $B_s^{\ast}\pi$     &$D_s^{\ast}\pi$ &$\frac{3}{2}(1^{+})$   &$B^{\ast}\pi$, $B\rho$ & $D^{\ast}\pi$, $D\rho$  \\
$0(2^{+})$   & $-$     &$D_s^{\ast}\omega$ &$\frac{1}{2}(2^{+})$   &$-$ &  $D^{\ast}(2007)^0\omega/\rho$ \\
$1(2^{+})$   & $-$     &$D_s^{\ast}\rho$ &$\frac{3}{2}(2^{+})$   &$-$ & $D^{\ast}(2007)^0\rho$  \\
$0(0^{-})$   &  $B_s^0\sigma$ &$D_s\sigma/f_0(980)$ &$\frac{1}{2}(0^{-})$   &$B\sigma$ & $D\sigma/a_0/f_0$, $D_{0}^\ast(2400)\pi$   \\
$1(0^{-})$   &  $-$     &$D_s a_0(980)$, $D_{s0}^\ast(2317)\pi$ &$\frac{3}{2}(0^{-})$   &$-$  &$Da_0(980)$, $D_{0}^\ast(2400)\pi$  \\
$0(1^{-})$   &  $-$     &$D_sh_1(1170)/f_1(1285)$, &$\frac{1}{2}(1^{-})$   &$B_{1}(5721)^0\pi$  & $Dh_1/f_1/a_1/b_1$, $D_{0}^\ast\omega/\rho$, \\
& &$D_{s}^{\ast}\sigma/f_0(980)$, $D_{s0}^\ast(2317)\omega$, &&&$D^{\ast}\sigma/f_0/a_0$, $D^{\ast}h_1/b_1$,  \\
& &$D_{s1}(2536)\omega$, $D_{s}^{\ast}h_1(1170)$  &&& $D_{1}(2420)^0\omega/\rho/\pi$  \\
$1(1^{-})$   &  $B_{s1}(5830)^0\pi$   &$D_sb_1(1235)/a_1(1260)$, &$\frac{3}{2}(1^{-})$   &$B_{1}(5721)^0\pi$  & $D^{\ast}a_0$, $Da_1/b_1$, $D_{0}^\ast\rho$, \\
& &$D_{s1}(2460)\pi$, $D_{s0}^\ast(2317)\rho$,   &&&$D_{0}^\ast(2400)\pi/\rho$,   \\
& &$D_{s}^{\ast}a_0(980)$, $D_{s1}(2536)\rho$   &&&$D^{\ast}(2007)^0b_1$  \\
\hlinewd{.8pt}
\end{tabular*}
\caption{Possible S-wave decay modes for the open-flavor charmed/bottom tetraquark states where `$-$' denotes that the predicted tetraquark masses are below all allowed S-wave two-body hadronic decay thresholds. \label{S-wavedecays}}
\end{table}
\begin{table}
\renewcommand{\arraystretch}{1.5}
\begin{tabular*}{17.5cm}{ccc|ccc}
\hlinewd{.8pt}
$I(J^{P})$  ~~   &~~ $sq\bar q\bar b$    ~&~ $sq\bar q\bar c$~ & ~$I(J^{P})$  ~~   & $qq\bar q\bar b$  ~ & ~$qq\bar q\bar c$~  \\
\hline
$0(0^{+})$   & $B_s^0h_1/f_1$, $B_s^{\ast}\sigma/f_0$, $BK_1(1270)$,
& $D_s^\ast\sigma$ &$\frac{1}{2}(0^{+})$   &$Bh_1/f_1/b_1/a_1$,  & $D_1(2420)\pi$, $D^\ast\sigma$,  \\
& $B_{s1}(5830)^0\omega/\phi$, $B_s^{\ast}h_1/f_1$, & &&$B^{\ast}\sigma/f_0/h_1/b_1/a_1$,&\\
& $B_{1}(5721)^0K^\ast(892)$, $B^\ast K_1/K_0^\ast$&&&$B_1(5721)^0\pi/\rho/\omega/\phi$&\\
$1(0^{+})$   & $B_{s1}(5830)^0\pi/\rho$, $B_s^{\ast}a_0$, $BK_1$, & $D_{s1}(2460)\pi$&$\frac{3}{2}(0^{+})$   &$Bb_1/a_1$,  & $D_1(2420)\pi$,  \\
& $B_s^0b_1/a_1$, $B_s^\ast a_0/b_1/a_1$, & &&$B^{\ast}a_0/b_1/a_1$,&\\
& $B_{1}(5721)^0K^\ast(892)$, $B^\ast K_1/K_0^\ast$ &&&$B_1(5721)^0\pi/\rho$&\\
$0(1^{+})$   & $B_s^0/B_s^\ast\sigma$    &$D_s/D_s^\ast\sigma$&$\frac{1}{2}(1^{+})$   &$B/B^{\ast}\sigma$, $B_{1}(5721)^0\pi$ & $D_1(2420)\pi$, $D/D^\ast\sigma$  \\
$1(1^{+})$   & $B_{s1}(5830)^0\pi$     &$D_{s1}(2460)\pi$ &$\frac{3}{2}(1^{+})$   &$B_{1}(5721)^0\pi$ & $D_1(2420)\pi$  \\
$0(2^{+})$   & $B_s^\ast\sigma$     &$D_s^{\ast}\sigma$ &$\frac{1}{2}(2^{+})$   &$B^{\ast}\sigma$, $B_{1}(5721)^0\pi$ &  $D_1(2420)\pi$, $D^\ast\sigma$ \\
$1(2^{+})$   &  $B_{s1}(5830)^0\pi$     &$D_{s1}(2460)\pi$ &$\frac{3}{2}(2^{+})$   &$B_{1}(5721)^0\pi$ & $D_1(2420)\pi$  \\
$0(0^{-})$   &  $B_s^0\omega$, $B_s^\ast\eta/\eta^\prime/\omega$, &$D_s\omega$, $D_s^\ast\eta/\eta^\prime\omega$, $D_{s1}\sigma$, &$\frac{1}{2}(0^{-})$   &$B^\ast\pi/\eta/\eta^\prime$, & $D\rho/\omega/\phi$, $D^\ast\pi/\eta/\eta^\prime$,   \\
& $B^\ast K$, $B/B^\ast K^\ast$  &$D^\ast K$, $D/D^\ast K^\ast$ &   &$B\rho/\omega$, $B^\ast\rho/\omega$ & $D^\ast\rho/\omega/\phi$, $D_1(2420)\sigma$   \\
$1(0^{-})$   &  $B_s^0\rho$, $B_s^\ast\pi/\rho$,  $B^\ast K$, $B/B^\ast K^\ast$   &$D_s\rho$, $D_s^\ast\pi/\rho$,  $D^\ast K$, $D/D^\ast K^\ast$ &$\frac{3}{2}(0^{-})$   &$B^\ast\pi/\rho, B\rho$  &$D/D^\ast\rho$, $D^\ast\pi$  \\
$0(1^{-})$   &  $B_s^0\omega, BK$,     &$D_s\eta/\eta^\prime/\omega$, $D_s^\ast\omega/\eta/\eta^\prime$, $D_{s1}\sigma$, &$\frac{1}{2}(1^{-})$   &$B/B^\ast\pi$  & $D\pi/\eta/\eta^\prime/\rho/\omega/\phi$, \\
& &$D/D^\ast K$, $D/D^\ast K^\ast$ &&&$D^\ast\pi/\eta/\eta^\prime/\rho/\omega/\phi$  \\
$1(1^{-})$   & $B_s^\ast\pi$, $B_s^0\rho/\pi, BK$    &$D_s\pi/\rho$, $D_s^\ast\pi/\rho$, &$\frac{3}{2}(1^{-})$   &$B/B^\ast\pi$   & $D\pi/\rho$, $D^\ast\pi/\rho$  \\
& &$D/D^\ast K$, $D/D^\ast K^\ast$   &&&   \\
\hlinewd{.8pt}
\end{tabular*}
\caption{Possible P-wave decay modes for the open-flavor charmed/bottom tetraquark states. \label{P-wavedecays}}
\end{table}

\section{Conclusions and discussion}\label{Sec:conclusion}

In this paper, we have studied the open-flavor charmed/bottom $sq\bar q\bar c$, $qq\bar q\bar c$, $sq\bar q\bar b$,
$qq\bar q\bar b$ tetraquark states with the spin-parity quantum numbers $J^P=0^+, 1^+, 2^+$ and $0^-, 1^-$. In the
diquark configurations, we use only the color-antisymmetric scalar and axial-vector diquarks to compose the color-antisymmetric $[\mathbf{\bar 3_c}]_{diquark} \otimes [\mathbf{3_c}]_{antidiquark}$ tetraquark interpolating currents.
Finally, we obtain five tetraquark currents in Eq.~\eqref{currents} with various spin-parity quantum numbers.

After performing the numerical analyses, we obtained the hadron masses for the open-flavor charmed/bottom
$sq\bar q\bar c$, $qq\bar q\bar c$, $sq\bar q\bar b$, $qq\bar q\bar b$ tetraquark states. For the charm-strange
$sq\bar q\bar c$ systems, we extract the hadron mass $m_{X_{cs}}=(2.55\pm0.12)$ GeV using
the interpolating current $J_{3\mu}(x)$ with $J^P=1^+$, which is not far from the mass of the $D_{s1}(2460)$ meson
within the error. In the scalar channel, however, the results for the $sq\bar q\bar c$ systems disfavor the tetraquark
explanation of the charm-strange $D_{s0}^\ast(2317)$ meson.

Our results indicate that many other charmed/bottom tetraquarks may exist, and we have evaluated their masses.
The tetraquarks $[su][\bar d\bar c]$, $\frac{[su][\bar u\bar c]+[sd][\bar d\bar c]}{\sqrt{2}}$, and $[sd][\bar u\bar c]$
can form an iso-triplet. Since we do not differentiate the up and down quarks in the OPE series, these tetraquark
states in the same isospin multiplet have the same extracted hadron masses in our analyses. In other words,
the mass spectra in Tables \ref{csudmass}--\ref{bqqqmass} contain all open-flavor charmed/bottom tetraquarks.
Among these states, the exotic doubly-charged tetraquarks, such as $[sd][\bar u\bar c]\to D_s^{(\ast)-}\pi^-$, is
especially interesting, and have not been observed so far. Our results for their mass spectra can
be useful for their searches in future experiments at facilities such as BESIII, BelleII, PANDA, LHCb, CMS, etc.

\section*{Acknowledgments}

This project is supported by the Natural Sciences and Engineering Research Council of
Canada (NSERC) and the National Natural Science Foundation of China under Grants
No. 11475015, No. 11375024, No. 11222547, No. 11175073, No. 11575008, and No. 11621131001;
the 973 program; the Ministry of Education of China (SRFDP under Grant No. 20120211110002 and the Fundamental Research
Funds for the Central Universities); the National Program for Support of Top-notch Youth Professionals.


\begin{thebibliography}{57}
\expandafter\ifx\csname natexlab\endcsname\relax\def\natexlab#1{#1}\fi
\expandafter\ifx\csname bibnamefont\endcsname\relax
  \def\bibnamefont#1{#1}\fi
\expandafter\ifx\csname bibfnamefont\endcsname\relax
  \def\bibfnamefont#1{#1}\fi
\expandafter\ifx\csname citenamefont\endcsname\relax
  \def\citenamefont#1{#1}\fi
\expandafter\ifx\csname url\endcsname\relax
  \def\url#1{\texttt{#1}}\fi
\expandafter\ifx\csname urlprefix\endcsname\relax\def\urlprefix{URL }\fi
\providecommand{\bibinfo}[2]{#2}
\providecommand{\eprint}[2][]{\url{#2}}

\bibitem[{\citenamefont{Patrignani
  et~al.}(2016)}]{2016-Patrignani-p100001-100001}
\bibinfo{author}{\bibfnamefont{C.}~\bibnamefont{Patrignani}}
  \bibnamefont{et~al.} (\bibinfo{collaboration}{Particle Data Group}),
  \bibinfo{journal}{Chin. Phys.} \textbf{\bibinfo{volume}{C40}},
  \bibinfo{pages}{100001} (\bibinfo{year}{2016}).

\bibitem[{\citenamefont{Aaij et~al.}(2015)}]{2015-Aaij-p72001-72001}
\bibinfo{author}{\bibfnamefont{R.}~\bibnamefont{Aaij}} \bibnamefont{et~al.}
  (\bibinfo{collaboration}{LHCb}), \bibinfo{journal}{Phys. Rev. Lett.}
  \textbf{\bibinfo{volume}{115}}, \bibinfo{pages}{072001}
  (\bibinfo{year}{2015}).

\bibitem[{\citenamefont{Klempt and Zaitsev}(2007)}]{2007-Klempt-p1-202}
\bibinfo{author}{\bibfnamefont{E.}~\bibnamefont{Klempt}} \bibnamefont{and}
  \bibinfo{author}{\bibfnamefont{A.}~\bibnamefont{Zaitsev}},
  \bibinfo{journal}{Phys. Rept.} \textbf{\bibinfo{volume}{454}},
  \bibinfo{pages}{1} (\bibinfo{year}{2007}).

\bibitem[{\citenamefont{Chen et~al.}(2016{\natexlab{a}})\citenamefont{Chen,
  Chen, Liu, and Zhu}}]{2016-Chen-p1-121}
\bibinfo{author}{\bibfnamefont{H.-X.} \bibnamefont{Chen}},
  \bibinfo{author}{\bibfnamefont{W.}~\bibnamefont{Chen}},
  \bibinfo{author}{\bibfnamefont{X.}~\bibnamefont{Liu}}, \bibnamefont{and}
  \bibinfo{author}{\bibfnamefont{S.-L.} \bibnamefont{Zhu}},
  \bibinfo{journal}{Phys. Rept.} \textbf{\bibinfo{volume}{639}},
  \bibinfo{pages}{1} (\bibinfo{year}{2016}{\natexlab{a}}), \eprint{arXiv:1601.02092}.

\bibitem[{\citenamefont{Esposito et~al.}(2015)\citenamefont{Esposito,
  Guerrieri, Piccinini, Pilloni, and Polosa}}]{2015-Esposito-p1530002-1530002}
\bibinfo{author}{\bibfnamefont{A.}~\bibnamefont{Esposito}},
  \bibinfo{author}{\bibfnamefont{A.~L.} \bibnamefont{Guerrieri}},
  \bibinfo{author}{\bibfnamefont{F.}~\bibnamefont{Piccinini}},
  \bibinfo{author}{\bibfnamefont{A.}~\bibnamefont{Pilloni}}, \bibnamefont{and}
  \bibinfo{author}{\bibfnamefont{A.~D.} \bibnamefont{Polosa}},
  \bibinfo{journal}{Int.J.Mod.Phys.} \textbf{\bibinfo{volume}{A30}},
  \bibinfo{pages}{1530002} (\bibinfo{year}{2015}).

\bibitem[{\citenamefont{Olsen}(2015)}]{2015-Olsen-p101401-101401}
\bibinfo{author}{\bibfnamefont{S.~L.} \bibnamefont{Olsen}},
  \bibinfo{journal}{Front.Phys.} \textbf{\bibinfo{volume}{10}},
  \bibinfo{pages}{101401} (\bibinfo{year}{2015}).

\bibitem[{\citenamefont{Lebed et~al.}(2016)\citenamefont{Lebed, Mitchell, and
  Swanson}}]{2016-Lebed-p-}
\bibinfo{author}{\bibfnamefont{R.~F.} \bibnamefont{Lebed}},
  \bibinfo{author}{\bibfnamefont{R.~E.} \bibnamefont{Mitchell}},
  \bibnamefont{and} \bibinfo{author}{\bibfnamefont{E.~S.}
  \bibnamefont{Swanson}} (\bibinfo{year}{2016}), \eprint{arXiv:1610.04528}.

\bibitem[{\citenamefont{Abazov et~al.}(2016)}]{2016-Abazov-p22003-22003}
\bibinfo{author}{\bibfnamefont{V.~M.} \bibnamefont{Abazov}}
  \bibnamefont{et~al.} (\bibinfo{collaboration}{D0}), \bibinfo{journal}{Phys.
  Rev. Lett.} \textbf{\bibinfo{volume}{117}}, \bibinfo{pages}{022003}
  (\bibinfo{year}{2016}).

\bibitem[{\citenamefont{Aaij et~al.}(2016)}]{2016-Aaij-p152003-152003}
\bibinfo{author}{\bibfnamefont{R.}~\bibnamefont{Aaij}} \bibnamefont{et~al.}
  (\bibinfo{collaboration}{LHCb}), \bibinfo{journal}{Phys. Rev. Lett.}
  \textbf{\bibinfo{volume}{117}}, \bibinfo{pages}{152003}
  (\bibinfo{year}{2016}).

\bibitem[{\citenamefont{Collaboration}(2016)}]{2016-Collaboration-p-a}
\bibinfo{author}{\bibfnamefont{The CMS Collaboration}},
  \bibinfo{collaboration}{CMS-PAS-BPH-16-002} (\bibinfo{year}{2016}).

  \bibitem[{\citenamefont{Collaboration}(2016)}]{D0:X5568}
\bibinfo{author}{\bibfnamefont{The D0 Collaboration}},
  \bibinfo{collaboration}{http://indico.cern.ch/event/432527/contributions/1072024/}(\bibinfo{year}{2016}).

\bibitem[{\citenamefont{Chen et~al.}(2016{\natexlab{b}})\citenamefont{Chen,
  Chen, Liu, Steele, and Zhu}}]{2016-Chen-p22002-22002}
\bibinfo{author}{\bibfnamefont{W.}~\bibnamefont{Chen}},
  \bibinfo{author}{\bibfnamefont{H.-X.} \bibnamefont{Chen}},
  \bibinfo{author}{\bibfnamefont{X.}~\bibnamefont{Liu}},
  \bibinfo{author}{\bibfnamefont{T.~G.} \bibnamefont{Steele}},
  \bibnamefont{and} \bibinfo{author}{\bibfnamefont{S.-L.} \bibnamefont{Zhu}},
  \bibinfo{journal}{Phys. Rev. Lett.} \textbf{\bibinfo{volume}{117}},
  \bibinfo{pages}{022002} (\bibinfo{year}{2016}{\natexlab{b}}),
  \eprint{arXiv:1602.08916}.

\bibitem[{\citenamefont{Agaev et~al.}(2016{\natexlab{a}})\citenamefont{Agaev,
  Azizi, and Sundu}}]{2016-Agaev-p74024-74024}
\bibinfo{author}{\bibfnamefont{S.~S.} \bibnamefont{Agaev}},
  \bibinfo{author}{\bibfnamefont{K.}~\bibnamefont{Azizi}}, \bibnamefont{and}
  \bibinfo{author}{\bibfnamefont{H.}~\bibnamefont{Sundu}},
  \bibinfo{journal}{Phys. Rev.} \textbf{\bibinfo{volume}{D93}},
  \bibinfo{pages}{074024} (\bibinfo{year}{2016}{\natexlab{a}}).

\bibitem[{\citenamefont{Zanetti et~al.}(2016)\citenamefont{Zanetti, Nielsen,
  and Khemchandani}}]{2016-Zanetti-p96011-96011}
\bibinfo{author}{\bibfnamefont{C.~M.} \bibnamefont{Zanetti}},
  \bibinfo{author}{\bibfnamefont{M.}~\bibnamefont{Nielsen}}, \bibnamefont{and}
  \bibinfo{author}{\bibfnamefont{K.~P.} \bibnamefont{Khemchandani}},
  \bibinfo{journal}{Phys. Rev.} \textbf{\bibinfo{volume}{D93}},
  \bibinfo{pages}{096011} (\bibinfo{year}{2016}).

\bibitem[{\citenamefont{Wang}(2016{\natexlab{a}})}]{2016-Wang-p335-339}
\bibinfo{author}{\bibfnamefont{Z.-G.} \bibnamefont{Wang}},
  \bibinfo{journal}{Commun. Theor. Phys.} \textbf{\bibinfo{volume}{66}},
  \bibinfo{pages}{335} (\bibinfo{year}{2016}{\natexlab{a}}).

\bibitem[{\citenamefont{Wang}(2016{\natexlab{b}})}]{2016-Wang-p279-279}
\bibinfo{author}{\bibfnamefont{Z.-G.} \bibnamefont{Wang}},
  \bibinfo{journal}{Eur. Phys. J.} \textbf{\bibinfo{volume}{C76}},
  \bibinfo{pages}{279} (\bibinfo{year}{2016}{\natexlab{b}}).

\bibitem[{\citenamefont{Wang and Zhu}(2016)}]{2016-Wang-p93101-93101}
\bibinfo{author}{\bibfnamefont{W.}~\bibnamefont{Wang}} \bibnamefont{and}
  \bibinfo{author}{\bibfnamefont{R.}~\bibnamefont{Zhu}},
  \bibinfo{journal}{Chin. Phys.} \textbf{\bibinfo{volume}{C40}},
  \bibinfo{pages}{093101} (\bibinfo{year}{2016}).

\bibitem[{\citenamefont{Tang and Qiao}(2016)}]{2016-Tang-p558-558}
\bibinfo{author}{\bibfnamefont{L.}~\bibnamefont{Tang}} \bibnamefont{and}
  \bibinfo{author}{\bibfnamefont{C.-F.} \bibnamefont{Qiao}},
  \bibinfo{journal}{Eur. Phys. J.} \textbf{\bibinfo{volume}{C76}},
  \bibinfo{pages}{558} (\bibinfo{year}{2016}).

\bibitem[{\citenamefont{Agaev et~al.}(2016{\natexlab{b}})\citenamefont{Agaev,
  Azizi, and Sundu}}]{2016-Agaev-p114007-114007}
\bibinfo{author}{\bibfnamefont{S.~S.} \bibnamefont{Agaev}},
  \bibinfo{author}{\bibfnamefont{K.}~\bibnamefont{Azizi}}, \bibnamefont{and}
  \bibinfo{author}{\bibfnamefont{H.}~\bibnamefont{Sundu}},
  \bibinfo{journal}{Phys. Rev.} \textbf{\bibinfo{volume}{D93}},
  \bibinfo{pages}{114007} (\bibinfo{year}{2016}{\natexlab{b}}).

\bibitem[{\citenamefont{Dias et~al.}(2016)\citenamefont{Dias, Khemchandani,
  Martínez~Torres, Nielsen, and Zanetti}}]{2016-Dias-p235-238}
\bibinfo{author}{\bibfnamefont{J.~M.} \bibnamefont{Dias}},
  \bibinfo{author}{\bibfnamefont{K.~P.} \bibnamefont{Khemchandani}},
  \bibinfo{author}{\bibfnamefont{A.}~\bibnamefont{Martínez~Torres}},
  \bibinfo{author}{\bibfnamefont{M.}~\bibnamefont{Nielsen}}, \bibnamefont{and}
  \bibinfo{author}{\bibfnamefont{C.~M.} \bibnamefont{Zanetti}},
  \bibinfo{journal}{Phys. Lett.} \textbf{\bibinfo{volume}{B758}},
  \bibinfo{pages}{235} (\bibinfo{year}{2016}).

\bibitem[{\citenamefont{Albuquerque et~al.}(2016)\citenamefont{Albuquerque,
  Narison, Rabemananjara, and
  Rabetiarivony}}]{2016-Albuquerque-p1650093-1650093}
\bibinfo{author}{\bibfnamefont{R.}~\bibnamefont{Albuquerque}},
  \bibinfo{author}{\bibfnamefont{S.}~\bibnamefont{Narison}},
  \bibinfo{author}{\bibfnamefont{A.}~\bibnamefont{Rabemananjara}},
  \bibnamefont{and}
  \bibinfo{author}{\bibfnamefont{D.}~\bibnamefont{Rabetiarivony}},
  \bibinfo{journal}{Int. J. Mod. Phys.} \textbf{\bibinfo{volume}{A31}},
  \bibinfo{pages}{1650093} (\bibinfo{year}{2016}).

\bibitem[{\citenamefont{Liu et~al.}(2016)\citenamefont{Liu, Liu, and
  Zhu}}]{2016-Liu-p74023-74023}
\bibinfo{author}{\bibfnamefont{Y.-R.} \bibnamefont{Liu}},
  \bibinfo{author}{\bibfnamefont{X.}~\bibnamefont{Liu}}, \bibnamefont{and}
  \bibinfo{author}{\bibfnamefont{S.-L.} \bibnamefont{Zhu}},
  \bibinfo{journal}{Phys. Rev.} \textbf{\bibinfo{volume}{D93}},
  \bibinfo{pages}{074023} (\bibinfo{year}{2016}).

\bibitem[{\citenamefont{Stancu}(2016)}]{2016-Stancu-p105001-105001}
\bibinfo{author}{\bibfnamefont{F.}~\bibnamefont{Stancu}}, \bibinfo{journal}{J.
  Phys.} \textbf{\bibinfo{volume}{G43}}, \bibinfo{pages}{105001}
  (\bibinfo{year}{2016}).

\bibitem[{\citenamefont{Ali et~al.}(2016)\citenamefont{Ali, Maiani, Polosa, and
  Riquer}}]{2016-Ali-p34036-34036}
\bibinfo{author}{\bibfnamefont{A.}~\bibnamefont{Ali}},
  \bibinfo{author}{\bibfnamefont{L.}~\bibnamefont{Maiani}},
  \bibinfo{author}{\bibfnamefont{A.~D.} \bibnamefont{Polosa}},
  \bibnamefont{and} \bibinfo{author}{\bibfnamefont{V.}~\bibnamefont{Riquer}},
  \bibinfo{journal}{Phys. Rev.} \textbf{\bibinfo{volume}{D94}},
  \bibinfo{pages}{034036} (\bibinfo{year}{2016}).

\bibitem[{\citenamefont{He and Ko}(2016)}]{2016-He-p92-97}
\bibinfo{author}{\bibfnamefont{X.-G.} \bibnamefont{He}} \bibnamefont{and}
  \bibinfo{author}{\bibfnamefont{P.}~\bibnamefont{Ko}}, \bibinfo{journal}{Phys.
  Lett.} \textbf{\bibinfo{volume}{B761}}, \bibinfo{pages}{92}
  (\bibinfo{year}{2016}).

\bibitem[{\citenamefont{Burns and Swanson}(2016)}]{2016-Burns-p627-633}
\bibinfo{author}{\bibfnamefont{T.~J.} \bibnamefont{Burns}} \bibnamefont{and}
  \bibinfo{author}{\bibfnamefont{E.~S.} \bibnamefont{Swanson}},
  \bibinfo{journal}{Phys. Lett.} \textbf{\bibinfo{volume}{B760}},
  \bibinfo{pages}{627} (\bibinfo{year}{2016}).

\bibitem[{\citenamefont{Guo et~al.}(2016)\citenamefont{Guo, Meißner, and
  Zou}}]{2016-Guo-p593-595}
\bibinfo{author}{\bibfnamefont{F.-K.} \bibnamefont{Guo}},
  \bibinfo{author}{\bibfnamefont{U.-G.} \bibnamefont{Meißner}},
  \bibnamefont{and} \bibinfo{author}{\bibfnamefont{B.-S.} \bibnamefont{Zou}},
  \bibinfo{journal}{Commun. Theor. Phys.} \textbf{\bibinfo{volume}{65}},
  \bibinfo{pages}{593} (\bibinfo{year}{2016}).

\bibitem[{\citenamefont{Lü and Dong}(2016)}]{2016-Lue-p-}
\bibinfo{author}{\bibfnamefont{Q.-F.} \bibnamefont{Lü}} \bibnamefont{and}
  \bibinfo{author}{\bibfnamefont{Y.-B.} \bibnamefont{Dong}}
  (\bibinfo{year}{2016}), \eprint{arXiv:1603.06417}.

\bibitem[{\citenamefont{Goerke et~al.}(2016)\citenamefont{Goerke, Gutsche,
  Ivanov, Korner, Lyubovitskij, and Santorelli}}]{2016-Goerke-p-}
\bibinfo{author}{\bibfnamefont{F.}~\bibnamefont{Goerke}},
  \bibinfo{author}{\bibfnamefont{T.}~\bibnamefont{Gutsche}},
  \bibinfo{author}{\bibfnamefont{M.~A.} \bibnamefont{Ivanov}},
  \bibinfo{author}{\bibfnamefont{J.~G.} \bibnamefont{Korner}},
  \bibinfo{author}{\bibfnamefont{V.~E.} \bibnamefont{Lyubovitskij}},
  \bibnamefont{and}
  \bibinfo{author}{\bibfnamefont{P.}~\bibnamefont{Santorelli}}
  (\bibinfo{year}{2016}), \eprint{arXiv:1608.04656}.

\bibitem[{\citenamefont{Agamaliev et~al.}(2016)\citenamefont{Agamaliev, Aliev,
  and Savcı}}]{2016-Agamaliev-p-}
\bibinfo{author}{\bibfnamefont{A.~K.} \bibnamefont{Agamaliev}},
  \bibinfo{author}{\bibfnamefont{T.~M.} \bibnamefont{Aliev}}, \bibnamefont{and}
  \bibinfo{author}{\bibfnamefont{M.}~\bibnamefont{Savcı}}
  (\bibinfo{year}{2016}), \eprint{arXiv:1610.03980}.

\bibitem[{\citenamefont{Agaev et~al.}(2016{\natexlab{c}})\citenamefont{Agaev,
  Azizi, and Sundu}}]{2016-Agaev-p351-351}
\bibinfo{author}{\bibfnamefont{S.~S.} \bibnamefont{Agaev}},
  \bibinfo{author}{\bibfnamefont{K.}~\bibnamefont{Azizi}}, \bibnamefont{and}
  \bibinfo{author}{\bibfnamefont{H.}~\bibnamefont{Sundu}},
  \bibinfo{journal}{Eur. Phys. J. Plus} \textbf{\bibinfo{volume}{131}},
  \bibinfo{pages}{351} (\bibinfo{year}{2016}{\natexlab{c}}).

\bibitem[{\citenamefont{Kang and Oller}(2016)}]{2016-Kang-p54010-54010}
\bibinfo{author}{\bibfnamefont{X.-W.} \bibnamefont{Kang}} \bibnamefont{and}
  \bibinfo{author}{\bibfnamefont{J.~A.} \bibnamefont{Oller}},
  \bibinfo{journal}{Phys. Rev.} \textbf{\bibinfo{volume}{D94}},
  \bibinfo{pages}{054010} (\bibinfo{year}{2016}).

\bibitem[{\citenamefont{Chen and Ping}(2016)}]{2016-Chen-p351-351}
\bibinfo{author}{\bibfnamefont{X.}~\bibnamefont{Chen}} \bibnamefont{and}
  \bibinfo{author}{\bibfnamefont{J.}~\bibnamefont{Ping}},
  \bibinfo{journal}{Eur. Phys. J.} \textbf{\bibinfo{volume}{C76}},
  \bibinfo{pages}{351} (\bibinfo{year}{2016}).

\bibitem[{\citenamefont{Albaladejo et~al.}(2016)\citenamefont{Albaladejo,
  Nieves, Oset, Sun, and Liu}}]{2016-Albaladejo-p515-519}
\bibinfo{author}{\bibfnamefont{M.}~\bibnamefont{Albaladejo}},
  \bibinfo{author}{\bibfnamefont{J.}~\bibnamefont{Nieves}},
  \bibinfo{author}{\bibfnamefont{E.}~\bibnamefont{Oset}},
  \bibinfo{author}{\bibfnamefont{Z.-F.} \bibnamefont{Sun}}, \bibnamefont{and}
  \bibinfo{author}{\bibfnamefont{X.}~\bibnamefont{Liu}},
  \bibinfo{journal}{Phys. Lett.} \textbf{\bibinfo{volume}{B757}},
  \bibinfo{pages}{515} (\bibinfo{year}{2016}).

\bibitem[{\citenamefont{Lang et~al.}(2016)\citenamefont{Lang, Mohler, and
  Prelovsek}}]{2016-Lang-p74509-74509}
\bibinfo{author}{\bibfnamefont{C.~B.} \bibnamefont{Lang}},
  \bibinfo{author}{\bibfnamefont{D.}~\bibnamefont{Mohler}}, \bibnamefont{and}
  \bibinfo{author}{\bibfnamefont{S.}~\bibnamefont{Prelovsek}},
  \bibinfo{journal}{Phys. Rev.} \textbf{\bibinfo{volume}{D94}},
  \bibinfo{pages}{074509} (\bibinfo{year}{2016}).

\bibitem[{\citenamefont{Chen and Liu}(2016)}]{2016-Chen-p34006-34006}
\bibinfo{author}{\bibfnamefont{R.}~\bibnamefont{Chen}} \bibnamefont{and}
  \bibinfo{author}{\bibfnamefont{X.}~\bibnamefont{Liu}},
  \bibinfo{journal}{Phys. Rev.} \textbf{\bibinfo{volume}{D94}},
  \bibinfo{pages}{034006} (\bibinfo{year}{2016}).

\bibitem[{\citenamefont{Lu et~al.}(2016)\citenamefont{Lu, Ren, and
  Geng}}]{2016-Lu-p-}
\bibinfo{author}{\bibfnamefont{J.-X.} \bibnamefont{Lu}},
  \bibinfo{author}{\bibfnamefont{X.-L.} \bibnamefont{Ren}}, \bibnamefont{and}
  \bibinfo{author}{\bibfnamefont{L.-S.} \bibnamefont{Geng}}
  (\bibinfo{year}{2016}), \eprint{arXiv:1607.06327}.

\bibitem[{\citenamefont{Sun et~al.}(2016)\citenamefont{Sun, Dong, and
  Pang}}]{2016-Sun-p-}
\bibinfo{author}{\bibfnamefont{B.-X.} \bibnamefont{Sun}},
  \bibinfo{author}{\bibfnamefont{F.-Y.} \bibnamefont{Dong}}, \bibnamefont{and}
  \bibinfo{author}{\bibfnamefont{J.-R.} \bibnamefont{Pang}}
  (\bibinfo{year}{2016}), \eprint{arXiv:1609.04068}.

\bibitem[{\citenamefont{Liu and Li}(2016)}]{2016-Liu-p455-455}
\bibinfo{author}{\bibfnamefont{X.-H.} \bibnamefont{Liu}} \bibnamefont{and}
  \bibinfo{author}{\bibfnamefont{G.}~\bibnamefont{Li}}, \bibinfo{journal}{Eur.
  Phys. J.} \textbf{\bibinfo{volume}{C76}}, \bibinfo{pages}{455}
  (\bibinfo{year}{2016}).

\bibitem[{\citenamefont{Esposito et~al.}(2016)\citenamefont{Esposito, Pilloni,
  and Polosa}}]{2016-Esposito-p292-295}
\bibinfo{author}{\bibfnamefont{A.}~\bibnamefont{Esposito}},
  \bibinfo{author}{\bibfnamefont{A.}~\bibnamefont{Pilloni}}, \bibnamefont{and}
  \bibinfo{author}{\bibfnamefont{A.~D.} \bibnamefont{Polosa}},
  \bibinfo{journal}{Phys. Lett.} \textbf{\bibinfo{volume}{B758}},
  \bibinfo{pages}{292} (\bibinfo{year}{2016}).

\bibitem[{\citenamefont{Chen et~al.}(2016{\natexlab{c}})\citenamefont{Chen,
  Chen, Liu, Liu, and Zhu}}]{2016-Chen-p-}
\bibinfo{author}{\bibfnamefont{H.-X.} \bibnamefont{Chen}},
  \bibinfo{author}{\bibfnamefont{W.}~\bibnamefont{Chen}},
  \bibinfo{author}{\bibfnamefont{X.}~\bibnamefont{Liu}},
  \bibinfo{author}{\bibfnamefont{Y.-R.} \bibnamefont{Liu}}, \bibnamefont{and}
  \bibinfo{author}{\bibfnamefont{S.-L.} \bibnamefont{Zhu}}
  (\bibinfo{year}{2016}{\natexlab{c}}), \eprint{arXiv:1609.08928}.

\bibitem[{\citenamefont{Aubert et~al.}(2003)}]{2003-Aubert-p242001-242001}
\bibinfo{author}{\bibfnamefont{B.}~\bibnamefont{Aubert}} \bibnamefont{et~al.}
  (\bibinfo{collaboration}{BaBar}), \bibinfo{journal}{Phys. Rev. Lett.}
  \textbf{\bibinfo{volume}{90}}, \bibinfo{pages}{242001}
  (\bibinfo{year}{2003}), \eprint{hep-ex/0304021}.

\bibitem[{\citenamefont{Besson et~al.}(2003)}]{2003-Besson-p32002-32002}
\bibinfo{author}{\bibfnamefont{D.}~\bibnamefont{Besson}} \bibnamefont{et~al.}
  (\bibinfo{collaboration}{CLEO}), \bibinfo{journal}{Phys. Rev.}
  \textbf{\bibinfo{volume}{D68}}, \bibinfo{pages}{032002}
  (\bibinfo{year}{2003}), \bibinfo{note}{[Erratum: Phys.
  Rev.D75,119908(2007)]}, \eprint{hep-ex/0305100}.

\bibitem[{\citenamefont{Godfrey and Kokoski}(1991)}]{1991-Godfrey-p1679-1687}
\bibinfo{author}{\bibfnamefont{S.}~\bibnamefont{Godfrey}} \bibnamefont{and}
  \bibinfo{author}{\bibfnamefont{R.}~\bibnamefont{Kokoski}},
  \bibinfo{journal}{Phys. Rev.} \textbf{\bibinfo{volume}{D43}},
  \bibinfo{pages}{1679} (\bibinfo{year}{1991}).

\bibitem[{\citenamefont{Barnes et~al.}(2003)\citenamefont{Barnes, Close, and
  Lipkin}}]{2003-Barnes-p54006-54006}
\bibinfo{author}{\bibfnamefont{T.}~\bibnamefont{Barnes}},
  \bibinfo{author}{\bibfnamefont{F.~E.} \bibnamefont{Close}}, \bibnamefont{and}
  \bibinfo{author}{\bibfnamefont{H.~J.} \bibnamefont{Lipkin}},
  \bibinfo{journal}{Phys. Rev.} \textbf{\bibinfo{volume}{D68}},
  \bibinfo{pages}{054006} (\bibinfo{year}{2003}), \eprint{hep-ph/0305025}.

\bibitem[{\citenamefont{Chen and Li}(2004)}]{2004-Chen-p232001-232001}
\bibinfo{author}{\bibfnamefont{Y.-Q.} \bibnamefont{Chen}} \bibnamefont{and}
  \bibinfo{author}{\bibfnamefont{X.-Q.} \bibnamefont{Li}},
  \bibinfo{journal}{Phys. Rev. Lett.} \textbf{\bibinfo{volume}{93}},
  \bibinfo{pages}{232001} (\bibinfo{year}{2004}), \eprint{hep-ph/0407062}.

\bibitem[{\citenamefont{Shifman et~al.}(1979)\citenamefont{Shifman, Vainshtein,
  and Zakharov}}]{1979-Shifman-p385-447}
\bibinfo{author}{\bibfnamefont{M.~A.} \bibnamefont{Shifman}},
  \bibinfo{author}{\bibfnamefont{A.~I.} \bibnamefont{Vainshtein}},
  \bibnamefont{and} \bibinfo{author}{\bibfnamefont{V.~I.}
  \bibnamefont{Zakharov}}, \bibinfo{journal}{Nucl. Phys.}
  \textbf{\bibinfo{volume}{B147}}, \bibinfo{pages}{385} (\bibinfo{year}{1979}).

\bibitem[{\citenamefont{Reinders et~al.}(1985)\citenamefont{Reinders,
  Rubinstein, and Yazaki}}]{1985-Reinders-p1-1}
\bibinfo{author}{\bibfnamefont{L.~J.} \bibnamefont{Reinders}},
  \bibinfo{author}{\bibfnamefont{H.}~\bibnamefont{Rubinstein}},
  \bibnamefont{and} \bibinfo{author}{\bibfnamefont{S.}~\bibnamefont{Yazaki}},
  \bibinfo{journal}{Phys. Rept.} \textbf{\bibinfo{volume}{127}},
  \bibinfo{pages}{1} (\bibinfo{year}{1985}).

\bibitem[{\citenamefont{Colangelo and
  Khodjamirian}(2000)}]{2000-Colangelo-p1495-1576}
\bibinfo{author}{\bibfnamefont{P.}~\bibnamefont{Colangelo}} \bibnamefont{and}
  \bibinfo{author}{\bibfnamefont{A.}~\bibnamefont{Khodjamirian}},
  \bibinfo{journal}{Frontier of Particle Physics} \textbf{\bibinfo{volume}{3}}
  (\bibinfo{year}{2000}), \eprint{hep-ph/0010175}.

\bibitem[{\citenamefont{Jaffe}(2005)}]{2005-Jaffe-p1-45}
\bibinfo{author}{\bibfnamefont{R.}~\bibnamefont{Jaffe}},
  \bibinfo{journal}{Phys.Rept.} \textbf{\bibinfo{volume}{409}},
  \bibinfo{pages}{1} (\bibinfo{year}{2005}), \eprint{hep-ph/0409065}.

\bibitem[{\citenamefont{Du et~al.}(2013)\citenamefont{Du, Chen, Chen, and
  Zhu}}]{2013-Du-p14003-14003}
\bibinfo{author}{\bibfnamefont{M.-L.} \bibnamefont{Du}},
  \bibinfo{author}{\bibfnamefont{W.}~\bibnamefont{Chen}},
  \bibinfo{author}{\bibfnamefont{X.-L.} \bibnamefont{Chen}}, \bibnamefont{and}
  \bibinfo{author}{\bibfnamefont{S.-L.} \bibnamefont{Zhu}},
  \bibinfo{journal}{Phys.Rev.} \textbf{\bibinfo{volume}{D87}},
  \bibinfo{pages}{014003} (\bibinfo{year}{2013}).

\bibitem[{\citenamefont{Chen et~al.}(2014)\citenamefont{Chen, Steele, and
  Zhu}}]{2014-Chen-p54037-54037}
\bibinfo{author}{\bibfnamefont{W.}~\bibnamefont{Chen}},
  \bibinfo{author}{\bibfnamefont{T.}~\bibnamefont{Steele}}, \bibnamefont{and}
  \bibinfo{author}{\bibfnamefont{S.-L.} \bibnamefont{Zhu}},
  \bibinfo{journal}{Phys.Rev.} \textbf{\bibinfo{volume}{D89}},
  \bibinfo{pages}{054037} (\bibinfo{year}{2014}).

\bibitem[{\citenamefont{Chen and Zhu}(2011)}]{2011-Chen-p34010-34010}
\bibinfo{author}{\bibfnamefont{W.}~\bibnamefont{Chen}} \bibnamefont{and}
  \bibinfo{author}{\bibfnamefont{S.-L.} \bibnamefont{Zhu}},
  \bibinfo{journal}{Phys. Rev.} \textbf{\bibinfo{volume}{D83}},
  \bibinfo{pages}{034010} (\bibinfo{year}{2011}).

\bibitem[{\citenamefont{Chen and Zhu}(2010)}]{2010-Chen-p105018-105018}
\bibinfo{author}{\bibfnamefont{W.}~\bibnamefont{Chen}} \bibnamefont{and}
  \bibinfo{author}{\bibfnamefont{S.-L.} \bibnamefont{Zhu}},
  \bibinfo{journal}{Phys.Rev.} \textbf{\bibinfo{volume}{D81}},
  \bibinfo{pages}{105018} (\bibinfo{year}{2010}).

\bibitem[{\citenamefont{Govaerts et~al.}(1987)\citenamefont{Govaerts, Reinders,
  Francken, Gonze, and Weyers}}]{1987-Govaerts-p674-674}
\bibinfo{author}{\bibfnamefont{J.}~\bibnamefont{Govaerts}},
  \bibinfo{author}{\bibfnamefont{L.~J.} \bibnamefont{Reinders}},
  \bibinfo{author}{\bibfnamefont{P.}~\bibnamefont{Francken}},
  \bibinfo{author}{\bibfnamefont{X.}~\bibnamefont{Gonze}}, \bibnamefont{and}
  \bibinfo{author}{\bibfnamefont{J.}~\bibnamefont{Weyers}},
  \bibinfo{journal}{Nucl. Phys.} \textbf{\bibinfo{volume}{B284}},
  \bibinfo{pages}{674} (\bibinfo{year}{1987}).

\bibitem[{\citenamefont{Narison}(2012)}]{2012-Narison-p259-263}
\bibinfo{author}{\bibfnamefont{S.}~\bibnamefont{Narison}},
  \bibinfo{journal}{Phys.Lett.} \textbf{\bibinfo{volume}{B707}},
  \bibinfo{pages}{259} (\bibinfo{year}{2012}).

\bibitem[{\citenamefont{Kuhn et~al.}(2007)\citenamefont{Kuhn, Steinhauser, and
  Sturm}}]{2007-Kuhn-p192-215}
\bibinfo{author}{\bibfnamefont{J.~H.} \bibnamefont{Kuhn}},
  \bibinfo{author}{\bibfnamefont{M.}~\bibnamefont{Steinhauser}},
  \bibnamefont{and} \bibinfo{author}{\bibfnamefont{C.}~\bibnamefont{Sturm}},
  \bibinfo{journal}{Nucl.Phys.} \textbf{\bibinfo{volume}{B778}},
  \bibinfo{pages}{192} (\bibinfo{year}{2007}), \eprint{hep-ph/0702103}.

\end{thebibliography}

\appendix

\section{Spectral densities} \label{sec:rhos}
In this appendix, we collect the spectral densities for all interpolating currents defined in Eq.~\eqref{currents}.
To calculate these spectral densities, we use the momentum space propagators for the heavy quarks (bottom
and charm) and strange quark while coordinate space propagators for the light quarks
\begin{equation}
\begin{split}
iS_Q^{ab}&=\frac{i\delta^{ab}}{\hat{p}-m_Q}+\frac{i}{4}g_s\frac{\lambda^n_{ab}}{2}G^n_{\mu\nu}\frac{\sigma^{\mu\nu}(\hat{p}+m_Q)+
(\hat{p}+m_Q)\sigma^{\mu\nu}}{(p^2-m_Q^2)^2}
+\frac{i\delta^{ab}}{12}\langle g^2_s GG\rangle
m_Q\frac{p^2+m_Q\hat{p}}{(p^2-m_Q^2)^4}\, ,\\
iS_q^{ab}&=\frac{i\delta^{ab}}{2\pi^2x^4}\hat{x}+\frac{i}{32\pi^2}\frac{\lambda^n_{ab}}{2}g_sG^n_{\mu\nu}\frac{1}{x^2}(\sigma^{\mu\nu}\hat{x}
+\hat{x}\sigma^{\mu\nu})-\frac{\delta^{ab}}{12}\langle\bar{q}q\rangle+\frac{\delta^{ab}x^2}{192}\langle g_s\bar{q}\sigma \cdot Gq\rangle -\frac{m_q\delta^{ab}}{4\pi^2x^2}+\frac{i\delta^{ab}m_q\langle \bar{q}q\rangle}{48}\hat{x}\, ,
\end{split}
\end{equation}
where $\hat{x}=\gamma_\mu x^\mu$, $\hat{p}=\gamma_\mu p^\mu$.
The nonperturbative terms correlated to $\ss$ and $\sGsa$ are also calculated by considering the various strange quark
condensates. We will use the projectors defined in Eq.~\eqref{projectors} to pick out the different invariant functions and
also the spectral densities for the vector and tensor currents. Up to dimension eight, the spectral density can be written
as
\begin{equation}
\rho(s)=\rho^{pert}(s)+\rho^{\langle\bar{q}q\rangle}(s)+\rho^{\langle
GG\rangle}(s)+\rho^{\langle\bar{q}Gq\rangle}(s)+\rho^{\langle
\bar{q}q\rangle^2}(s)+\rho^{\qq\qGqb}\, .
\end{equation}

\begin{itemize}
\item For the current $J_1(x)$ with $J^P=0^+$
{\allowdisplaybreaks
\begin{align}
\nonumber
\rho^{pert}_1(s)&=\frac{1}{512\pi^6}\dab\frac{(1-\alpha-\beta)^2\f(s)^3(\alpha m_s^2+\beta m_Q^2-3\alpha\beta s)}{\alpha^3\beta^3}\, ,
\non
\rho^{\qq}_1(s)&=-\frac{\qq}{16\pi^4}\dab\left(\frac{m_Q}{\alpha}+\frac{m_s}{\beta}\right)\frac{(1-\alpha-\beta)\f(s)(\alpha m_s^2+\beta m_Q^2-2\alpha\beta s)}{\alpha\beta}\, ,
\non
\rho^{\ss}_1(s)&=\frac{m_s\ss(s-m_Q^2)(s^2-5m_Q^2s-2m_Q^4)}{384\pi^4s}-\frac{m_Q^4m_s\ss\log[m_Q^2/s]}{64\pi^4}\, ,
\non
\rho^{\GGa}_1(s)&=\frac{\GGb}{1024\pi^6}\dab\Bigg[
\left(\frac{m_Q^2}{\alpha^3}+\frac{m_s^2}{\beta^3}\right)\frac{(1-\alpha-\beta)^2(2\alpha m_s^2+2\beta m_Q^2-3\alpha\beta s)}{3}
\non &+\left(\frac{1}{\alpha}+\frac{1}{\beta}\right)\frac{(1-\alpha-\beta)\f(s)(\alpha m_s^2+\beta m_Q^2-2\alpha\beta s)}{\alpha\beta}\Bigg]\, ,
\\
\rho^{\qGqb}_1(s)&=\frac{\qGqa}{64\pi^4}\dab\Bigg[\left(\frac{m_Q}{\alpha}+\frac{m_s}{\beta}\right)
\non &
-\left(\frac{m_Q}{\alpha^2}+\frac{m_s}{\beta^2}\right)(1-\alpha-\beta)\Bigg](2\alpha m_s^2+2\beta m_Q^2-3\alpha\beta s)\, ,
\non
\rho^{\qq^2}_1(s)&=\frac{m_Qm_s\qq^2}{12\pi^2}\sqrt{\left(1-\frac{m_Q^2-m_s^2}{s}\right)^2-\frac{4m_s^2}{s}}
+\frac{\qq\ss(s-m_Q^2)(s-m_Qm_s-m_Q^2)}{24\pi^2s}\, ,
\non
\rho^{\qq\qGqb}_1(s)&=\frac{\ss\qGqa+\qq\sGsa}{48\pi^2}
\non &+\frac{m_Qm_s\qq\qGqa}{48\pi^2}\int_0^1d\alpha
\Bigg[\frac{2m_s^2}{\alpha^2}\delta'\left(s-\tilde{m}^2_Q\right)-
\frac{1}{\alpha(1-\alpha)}\delta\left(s-\tilde{m}^2_Q\right)
\Bigg]\, . \label{SD0+1}
\end{align}
}
in which
\begin{align}
\nonumber
\alpha_{max}&=\frac{1}{2}\left[1+\frac{m_Q^2-m_s^2}{s}+\sqrt{\left(1+\frac{m_Q^2-m_s^2}{s}\right)^2-\frac{4m_Q^2}{s}}\right]\, ,
\non
\alpha_{min}&=\frac{1}{2}\left[1+\frac{m_Q^2-m_s^2}{s}-\sqrt{\left(1+\frac{m_Q^2-m_s^2}{s}\right)^2-\frac{4m_Q^2}{s}}\right]\, ,
\\
\beta_{max}&=\frac{m_s^2\alpha}{s\alpha-m_Q^2}\, ,
\non
\beta_{min}&=1-\alpha\, ,
\non
\tilde{m}^2_Q&=\frac{m_Q^2\alpha+m_s^2(1-\alpha)}{\alpha(1-\alpha)}\, .
\end{align}

\item For the current $J_2(x)$ with $J^P=0^+$
{\allowdisplaybreaks
\begin{align}
\nonumber
\rho^{pert}_2(s)&=4\rho^{pert}_1(s), \, \rho^{\qq}_2(s)=8\rho^{\qq}_1(s), \, \rho^{\ss}_2(s)=4\rho^{\ss}_1(s)\, ,
\non
\rho^{\GGa}_2(s)&=\frac{\GGb}{256\pi^6}\dab\Bigg[
\left(\frac{m_Q^2}{\alpha^3}+\frac{m_s^2}{\beta^3}\right)\frac{(1-\alpha-\beta)^2(2\alpha m_s^2+2\beta m_Q^2-3\alpha\beta s)}{3}
\non &-\left(\frac{1}{\alpha}+\frac{1}{\beta}\right)\frac{(1-\alpha-\beta)\f(s)(\alpha m_s^2+\beta m_Q^2-2\alpha\beta s)}{\alpha\beta}\Bigg]\, ,
\\
\rho^{\qGqb}_2(s)&=\frac{\qGqa}{8\pi^4}\dab\left(\frac{m_Q}{\alpha}+\frac{m_s}{\beta}\right)(2\alpha m_s^2+2\beta m_Q^2-3\alpha\beta s)\, ,
\non
\rho^{\qq^2}_2(s)&=\frac{4m_Qm_s\qq^2}{3\pi^2}\sqrt{\left(1-\frac{m_Q^2-m_s^2}{s}\right)^2-\frac{4m_s^2}{s}}
+\frac{\qq\ss(s-m_Q^2)(s-m_Qm_s-m_Q^2)}{3\pi^2s}\, ,
\non
\rho^{\qq\qGqb}_2(s)&=\frac{\ss\qGqa+\qq\sGsa}{6\pi^2}+\frac{2m_Qm_s^3\qq\qGqa}{3\pi^2}\int_0^1d\alpha\frac{1}{\alpha^2}\delta'\left(s-\tilde{m}^2_Q\right)\, . \label{SD0+2}
\end{align}
}

\item For the trace of current $J_{5\mu\nu}(x)$ with $J^P=0^+$(T)
{\allowdisplaybreaks
\begin{align}
\nonumber
\rho^{pert}_3(s)&=\frac{1}{4}\rho^{pert}_1(s), \, \rho^{\qq}_3(s)=\frac{1}{8}\rho^{\qq}_1(s), \, \rho^{\ss}_3(s)=\frac{1}{4}\rho^{\ss}_1(s)\, ,
\non
\rho^{\GGa}_3(s)&=\frac{\GGb}{4096\pi^6}\dab\Bigg[
\frac{(1-\beta)\f(s)(\alpha m_s^2+\beta m_Q^2-2\alpha\beta s)}{4\alpha^2\beta}
\non &+\frac{m_Q^2(1-\alpha-\beta)^2(2\alpha m_s^2+2\beta m_Q^2-3\alpha\beta s)}{3\alpha^3}\Bigg]\, ,
\\
\rho^{\qGqb}_3(s)&=\frac{\qGqa}{2048\pi^4}\dab(2\alpha m_s^2+2\beta m_Q^2-3\alpha\beta s)
\non &
\Bigg[3\left(\frac{m_Q}{\alpha}+\frac{m_s}{\beta}\right)+\frac{(m_Q-m_s)(1-\alpha-\beta)}{\alpha\beta}\Bigg]\, ,
\non
\rho^{\qq^2}_3(s)&=\frac{m_Qm_s\qq^2}{48\pi^2}\sqrt{\left(1-\frac{m_Q^2-m_s^2}{s}\right)^2-\frac{4m_s^2}{s}}
+\frac{\qq\ss(s-m_Q^2)(s-m_Qm_s-m_Q^2)}{192\pi^2s}\, ,
\non
\rho^{\qq\qGqb}_3(s)&=\frac{\ss\qGqa}{384\pi^2}+\frac{\qq\sGsa}{768\pi^2}+\frac{m_Q^2\qq\sGsa}{1536\pi^2 s}
\non &+\frac{m_Qm_s\qq\qGqa}{768\pi^2}\int_0^1d\alpha
\Bigg[\frac{8m_s^2}{\alpha^2}\delta'\left(s-\tilde{m}^2_Q\right)-
\frac{1}{\alpha(1-\alpha)}\delta\left(s-\tilde{m}^2_Q\right)
\Bigg]\, . \label{SD0+T}
\end{align}
}

\item For the traceless symmetric part of the current $J_{5\mu\nu}(x)$ with $J^P=0^+$(S)
{\allowdisplaybreaks
\begin{align}
\nonumber
\rho^{pert}_4(s)&=\frac{1}{128\pi^6}\dab(1-\alpha-\beta)^2\f(s)^2\Bigg[\frac{(1-\alpha-\beta)(\alpha m_s^2+\beta m_Q^2-5\alpha\beta s)^2}{8\alpha^3\beta^3}
\non
&-\frac{(1-\alpha-\beta)(\alpha m_s^2+\beta m_Q^2-\alpha\beta s)^2s}{\alpha^2\beta^2}
-\frac{\f(s)(\alpha m_s^2+\beta m_Q^2-27\alpha\beta s)}{16\alpha^3\beta^3}\Bigg]\, ,
\non
\rho^{\qq}_4(s)&=-\frac{3\qq}{128\pi^4}\dab\left(\frac{m_Q}{\alpha}+\frac{m_s}{\beta}\right)\frac{(1-\alpha-\beta)\f(s)(\alpha m_s^2+\beta m_Q^2)}{\alpha\beta}\, ,
\non
\rho^{\ss}_4(s)&=\frac{m_s\ss(s-m_Q^2)(9s^4-45m_Q^2s^3-29m_Q^4s^2+7m_Q^6s-2m_Q^8)}{1536\pi^4s^3}-\frac{5m_Q^4m_s\ss\log[m_Q^2/s]}{128\pi^4}\, ,
\non
\rho^{\GGa}_4(s)&=\frac{\GGb}{1024\pi^6}\dab\Bigg\{
-\frac{(1-\alpha-\beta)\f(s)(17\alpha m_s^2+17\beta m_Q^2-46\alpha\beta s)}{48\alpha^2\beta}
\non &+\frac{m_Q^2(1-\alpha-\beta)^2(2\alpha m_s^2+2\beta m_Q^2-3\alpha\beta s)}{4\alpha^3}
-\frac{m_Q^2(1-\alpha-\beta)^3(\alpha m_s^2+\beta m_Q^2-3\alpha\beta s)}{3\alpha^3}
\non &+\frac{(1-\alpha-\beta)^2\left[9\f(s)^2-78\alpha\beta s\f(s)+28\alpha^2\beta^2 s^2\right]}{48\alpha^2\beta}\Bigg\}\, ,
\\
\rho^{\qGqb}_4(s)&=\frac{\qGqa}{2048\pi^4}\dab\Bigg[\frac{m_s(70\alpha m_s^2+70\beta m_Q^2-27\alpha\beta s)}{3\beta}
\non & -\frac{m_Q(1-\alpha-14\beta)(2\alpha m_s^2+2\beta m_Q^2-\alpha\beta s)}{\alpha\beta}+\frac{m_s(1-\alpha-\beta)(2\alpha m_s^2+2\beta m_Q^2-\alpha\beta s)}{\alpha\beta}\Bigg]\, ,
\non
\rho^{\qq^2}_4(s)&=\frac{m_Qm_s\qq^2}{16\pi^2}\sqrt{\left(1-\frac{m_Q^2-m_s^2}{s}\right)^2-\frac{4m_s^2}{s}}
\non &
+\frac{\qq\ss(s-m_Q^2)(s^2-3m_Qm_ss-5m_Q^2s+6m_Q^3m_s+4m_Q^4)}{192\pi^2s^2}\, ,
\non
\rho^{\qq\qGqb}_4(s)&=\frac{(14s^2-3m_Q^2s-24m_Q^4)\qq\sGsa}{1536\pi^2s^2}+\frac{(s^2-2m_Q^4)\ss\qGqa}{128\pi^2s^2}
\non &+\frac{m_Qm_s\qq\qGqa}{768\pi^2}\int_0^1d\alpha
\Bigg[\frac{24m_s^2}{\alpha^2}\delta'\left(s-\tilde{m}^2_Q\right)+
\frac{1}{\alpha(1-\alpha)}\delta\left(s-\tilde{m}^2_Q\right)
\Bigg]\, . \label{SD0+S}
\end{align}
}

\item For the current $J_{3\mu}(x)$ with $J^P=1^+$
{\allowdisplaybreaks
\begin{align}
\nonumber
\rho^{pert}_5(s)&=\frac{1}{1024\pi^6}\dab\frac{(1-\alpha-\beta)^2\f(s)^3(\alpha m_s^2+\beta m_Q^2-5\alpha\beta s)}{\alpha^3\beta^3}\, ,
\non
\rho^{\qq}_5(s)&=-\frac{\qq}{16\pi^4}\dab\Bigg[\frac{m_Q(1-\alpha-\beta)\f(s)(\alpha m_s^2+\beta m_Q^2-2\alpha\beta s)}{\alpha^2\beta}
\non &+\frac{m_s(1-\alpha-\beta)\f(s)(\alpha m_s^2+\beta m_Q^2-3\alpha\beta s)}{2\alpha\beta^2}\Bigg]\, ,
\non
\rho^{\ss}_5(s)&=\frac{m_s\ss(s-m_Q^2)(3s^3-13m_Q^2s^2-m_Q^4s-m_Q^6)}{1536\pi^4s^2}-\frac{m_Q^4m_s\ss\log[m_Q^2/s]}{128\pi^4}\, ,
\non
\rho^{\GGa}_5(s)&=\frac{\GGb}{3072\pi^6}\dab\Bigg\{
\left(\frac{m_Q^2}{\alpha^3}+\frac{m_s^2}{\beta^3}\right)(1-\alpha-\beta)^2(\alpha m_s^2+\beta m_Q^2-2\alpha\beta s)
\non &+\frac{(1-\alpha-\beta)\f(s)}{2\alpha\beta}\left[\frac{3(\alpha m_s^2+\beta m_Q^2-3\alpha\beta s)}{\beta}-\frac{(3\alpha m_s^2+3\beta m_Q^2-5\alpha\beta s)}{\alpha}\right]\Bigg\}\, ,
\\
\rho^{\qGqb}_5(s)&=\frac{\qGqa}{64\pi^4}\dab\Bigg[\frac{m_Q(2\alpha m_s^2+2\beta m_Q^2-3\alpha\beta s)}{\alpha}
\non &-\frac{m_s(1-\alpha-2\beta)(\alpha m_s^2+\beta m_Q^2-2\alpha\beta s)}{\beta^2}\Bigg]\, ,
\non
\rho^{\qq^2}_5(s)&=\frac{m_Qm_s\qq^2}{12\pi^2}\sqrt{\left(1-\frac{m_Q^2-m_s^2}{s}\right)^2-\frac{4m_s^2}{s}}
+\frac{\qq\ss(s-m_Q^2)(2s^2-3m_Qm_ss-m_Q^2s-m_Q^4)}{72\pi^2s^2}\, ,
\non
\rho^{\qq\qGqb}_5(s)&=\frac{(s^2+m_Q^4)(\ss\qGqa+\qq\sGsa)}{96\pi^2s^2}
\non &+\frac{m_Qm_s\qq\qGqa}{48\pi^2}\int_0^1d\alpha
\Bigg[\frac{2m_s^2}{\alpha^2}\delta'\left(s-\tilde{m}^2_Q\right)-
\frac{1}{\alpha}\delta\left(s-\tilde{m}^2_Q\right)
\Bigg]\, . \label{SD1+1}
\end{align}
}

\item For the current $J_{4\mu}(x)$ with $J^P=1^+$
{\allowdisplaybreaks
\begin{align}
\nonumber
\rho^{pert}_6(s)&=3\rho^{pert}_5(s), \, \rho^{\ss}_6(s)=3\rho^{\ss}_5(s)\, ,
\non
\rho^{\qq}_6(s)&=-\frac{3\qq}{16\pi^4}\dab\Bigg[\frac{m_Q(1-\alpha-\beta)\f(s)(\alpha m_s^2+\beta m_Q^2-3\alpha\beta s)}{2\alpha^2\beta}
\non &+\frac{m_s(1-\alpha-\beta)\f(s)(\alpha m_s^2+\beta m_Q^2-2\alpha\beta s)}{\alpha\beta^2}\Bigg]\, ,
\non
\rho^{\GGa}_6(s)&=-\frac{\GGb}{1024\pi^6}\dab\Bigg\{
\left(\frac{m_Q^2}{\alpha^3}+\frac{m_s^2}{\beta^3}\right)\frac{(1-\alpha-\beta)^2(\alpha m_s^2+\beta m_Q^2)}{3}
\non &+\frac{(1-\alpha-\beta)\f(s)}{2\alpha\beta}\left[\frac{3\alpha m_s^2+3\beta m_Q^2-5\alpha\beta s}{\beta}-\frac{3\alpha m_s^2+3\beta m_Q^2-9\alpha\beta s}{\alpha}\right]\Bigg\}\, ,
\\
\rho^{\qGqb}_6(s)&=-\frac{3\qGqa}{64\pi^4}\dab\Bigg[\frac{m_Q(1-2\alpha-\beta)(\alpha m_s^2+\beta m_Q^2-2\alpha\beta s)}{\alpha^2}
\non &+\frac{m_s(2\alpha m_s^2+2\beta m_Q^2-3\alpha\beta s)}{\beta}\Bigg]\, ,
\non
\rho^{\qq^2}_6(s)&=\frac{m_Qm_s\qq^2}{4\pi^2}\sqrt{\left(1-\frac{m_Q^2-m_s^2}{s}\right)^2-\frac{4m_s^2}{s}}
+\frac{\qq\ss(s-m_Q^2)(2s^2-m_Qm_ss-2m_Q^2s-m_Q^3m_s)}{16\pi^2s^2}\, ,
\non
\rho^{\qq\qGqb}_6(s)&=\frac{\ss\qGqa+\qq\sGsa}{16\pi^2}
\non &+\frac{m_Qm_s\qq\qGqa}{16\pi^2}\int_0^1d\alpha
\Bigg[\frac{2m_s^2}{\alpha^2}\delta'\left(s-\tilde{m}^2_Q\right)-
\frac{1}{\alpha}\delta\left(s-\tilde{m}^2_Q\right)
\Bigg]\, . \label{SD1+2}
\end{align}
}

\item For the traceless antisymmetric part of the current $J_{5\mu\nu}(x)$ with $J^P=1^+$(A)
{\allowdisplaybreaks
\begin{align}
\nonumber
\rho^{pert}_7(s)&=-\frac{3}{128\pi^6}\dab\frac{(1-\alpha-\beta)^2\f(s)^3s}{\alpha^2\beta^2}\, ,
\non
\rho^{\qq}_7(s)&=-\frac{3\qq}{16\pi^4}\dab\left(\frac{m_Q}{\alpha}+\frac{m_s}{\beta}\right)\frac{(1-\alpha-\beta)\f(s)(\alpha m_s^2+\beta m_Q^2-3\alpha\beta s)}{\alpha\beta}\, ,
\non
\rho^{\ss}_7(s)&=\frac{m_s\ss(s-m_Q^2)^4}{128\pi^4s^2}\, ,
\non
\rho^{\GGa}_7(s)&=\frac{\GGb}{512\pi^6}\dab\Bigg[
-\frac{(1-\alpha-\beta)\f(s)(3\alpha m_s^2+3\beta m_Q^2-5\alpha\beta s)}{4\alpha^2\beta}
\\ &+\frac{\f(s)(\alpha m_s^2+\beta m_Q^2-3\alpha\beta s)}{4\alpha\beta}
-\frac{m_Q^2(1-\alpha-\beta)^2\beta s}{\alpha^2}\Bigg]\, ,
\\
\rho^{\qGqb}_7(s)&=\frac{\qGqa}{128\pi^4}\dab\Bigg[11(\alpha m_s^2+\beta m_Q^2-2\alpha\beta s)\left(\frac{m_Q}{\alpha}+\frac{m_s}{\beta}\right)
\non & +\frac{(m_Q-m_s)(1-\alpha-\beta)(\alpha m_s^2+\beta m_Q^2-2\alpha\beta s)}{\alpha\beta}\Bigg]\, ,
\non
\rho^{\qq^2}_7(s)&=\frac{m_Qm_s\qq^2}{2\pi^2}\sqrt{\left(1-\frac{m_Q^2-m_s^2}{s}\right)^2-\frac{4m_s^2}{s}}
\non &
-\frac{\qq\ss(s-m_Q^2)(4s^2-3m_Qm_ss-2m_Q^2s-3m_Q^3m_s-2m_Q^4)}{24\pi^2s^2}\, ,
\non
\rho^{\qq\qGqb}_7(s)&=\frac{(5s^2+6m_Q^4)\qq\sGsa}{96\pi^2s^2}+\frac{(s^2+m_Q^4)\ss\qGqa}{16\pi^2s^2}
\non &+\frac{m_Qm_s\qq\qGqa}{96\pi^2}\int_0^1d\alpha
\Bigg[\frac{24m_s^2}{\alpha^2}\delta'\left(s-\tilde{m}^2_Q\right)-
\frac{1}{\alpha(1-\alpha)}\delta\left(s-\tilde{m}^2_Q\right)
\Bigg]\, . \label{SD1+A}
\end{align}
}

\item For the traceless antisymmetric part of the current $J_{5\mu\nu}(x)$ with $J^P=1^-$(A)
{\allowdisplaybreaks
\begin{align}
\nonumber
\rho^{pert}_8(s)&=\rho^{pert}_7(s), \, \rho^{\ss}_8(s)=\rho^{\ss}_7(s)\, ,
\non
\rho^{\qq}_8(s)&=\frac{3\qq}{16\pi^4}\dab\left(\frac{m_Q}{\alpha}+\frac{m_s}{\beta}\right)\frac{(1-\alpha-\beta)\f(s)^2}{\alpha\beta}\, ,
\non
\rho^{\GGa}_8(s)&=\frac{\GGb}{512\pi^6}\dab\Bigg[
\frac{(1-\alpha-\beta)\f(s)(3\alpha m_s^2+3\beta m_Q^2-7\alpha\beta s)}{4\alpha^2\beta}
\\ &-\frac{\f(s)^2}{4\alpha\beta}
-\frac{m_Q^2(1-\alpha-\beta)^2\beta s}{\alpha^2}\Bigg]\, ,
\\
\rho^{\qGqb}_8(s)&=-\frac{\qGqa}{128\pi^4}\dab\Bigg[11(\alpha m_s^2+\beta m_Q^2-2\alpha\beta s)\left(\frac{m_Q}{\alpha}+\frac{m_s}{\beta}\right)
\non & +\frac{(m_Q-m_s)(1-\alpha-\beta)(\alpha m_s^2+\beta m_Q^2-\alpha\beta s)}{\alpha\beta}\Bigg]\, ,
\non
\rho^{\qq^2}_8(s)&=-\frac{m_Qm_s\qq^2}{2\pi^2}\sqrt{\left(1-\frac{m_Q^2-m_s^2}{s}\right)^2-\frac{4m_s^2}{s}}
-\frac{\qq\ss(s-m_Q^2)^2(2s-3m_Qm_s-2m_Q^2)}{24\pi^2s^2}\, ,
\non
\rho^{\qq\qGqb}_8(s)&=-\frac{(s-m_Q^2)(5s+6m_Q^2)\qq\sGsa}{96\pi^2s^2}-\frac{(s^2-m_Q^4)\ss\qGqa}{16\pi^2s^2}
\non &-\frac{m_Qm_s\qq\qGqa}{96\pi^2}\int_0^1d\alpha
\Bigg[\frac{24m_s^2}{\alpha^2}\delta'\left(s-\tilde{m}^2_Q\right)-
\frac{1}{\alpha(1-\alpha)}\delta\left(s-\tilde{m}^2_Q\right)
\Bigg]\, . \label{SD1-A}
\end{align}
}

\item For the traceless symmetric part of the current $J_{5\mu\nu}(x)$ with $J^P=1^-$(S)
{\allowdisplaybreaks
\begin{align}
\nonumber
\rho^{pert}_9(s)&=-\frac{1}{64\pi^6}\dab\frac{(1-\alpha-\beta)^2\f(s)^2(\alpha m_s^2+\beta m_Q^2-3\alpha\beta s)}{\alpha^2\beta^2}
\non &\left[(1-\alpha-\beta)s+\frac{\f(s)}{4\alpha\beta}\right]\, ,
\non
\rho^{\qq}_9(s)&=\rho^{\qq}_8(s), \, \rho^{\qq^2}_9(s)=\rho^{\qq^2}_8(s)\, ,
\non
\rho^{\ss}_9(s)&=\frac{m_Q^2m_s\ss(s-m_Q^2)(3s-m_Q^2)(s^2+m_Q^2s+m_Q^4)}{192\pi^4s^3}+\frac{m_Q^4m_s\ss\log[m_Q^2/s]}{32\pi^4}\, ,
\non
\rho^{\GGa}_9(s)&=\frac{\GGb}{256\pi^6}\dab\Bigg[
\frac{(1-\alpha-\beta)\f(s)(5\alpha m_s^2+5\beta m_Q^2-9\alpha\beta s)}{8\alpha^2\beta}
\non &-\frac{m_Q^2(1-\alpha-\beta)^2\beta s}{2\alpha^2}-\frac{m_Q^2(1-\alpha-\beta)^3(\alpha m_s^2+\beta m_Q^2-3\alpha\beta s)}{3\alpha^3}
\non &+\frac{\f(s)^2}{8\alpha\beta}-\frac{(1-\alpha-\beta)^2\f(s)(\alpha m_s^2+\beta m_Q^2-5\alpha\beta s)}{4\alpha^2\beta}\Bigg]\, ,
\\
\rho^{\qGqb}_9(s)&=-\frac{\qGqa}{128\pi^4}\dab\Bigg[13(\alpha m_s^2+\beta m_Q^2-2\alpha\beta s)\left(\frac{m_Q}{\alpha}+\frac{m_s}{\beta}\right)
\non & -\frac{(m_Q-m_s)(1-\alpha-\beta)(\alpha m_s^2+\beta m_Q^2-\alpha\beta s)}{\alpha\beta}\Bigg]\, ,
\non
\rho^{\qq\qGqb}_9(s)&=-\frac{(s-m_Q^2)(7s+6m_Q^2)\qq\sGsa}{96\pi^2s^2}-\frac{(s^2-m_Q^4)\ss\qGqa}{16\pi^2s^2}
\non &-\frac{m_Qm_s\qq\qGqa}{96\pi^2}\int_0^1d\alpha
\Bigg[\frac{24m_s^2}{\alpha^2}\delta'\left(s-\tilde{m}^2_Q\right)+
\frac{1}{\alpha(1-\alpha)}\delta\left(s-\tilde{m}^2_Q\right)
\Bigg]\, . \label{SD1-S}
\end{align}
}

\item For the traceless symmetric part of the current $J_{5\mu\nu}(x)$ with $J^P=2^+$(S)
{\allowdisplaybreaks
\begin{align}
\nonumber
\rho^{pert}_{10}(s)&=\frac{1}{96\pi^6}\dab\frac{(1-\alpha-\beta)^2\f(s)^2}{\alpha^2\beta^2}
\non &\Bigg[\frac{(1-\alpha-\beta)(\alpha m_s^2+\beta m_Q^2-5\alpha\beta s)^2}{8\alpha\beta}+\frac{\f(s)(2\alpha m_s^2+2\beta m_Q^2-9\alpha\beta s)}{4\alpha\beta}
\non &-(1-\alpha-\beta)\f(s)s\Bigg]\, ,
\non
\rho^{\qq}_{10}(s)&=\frac{5}{3}\rho^{\qq}_7(s), \, \rho^{\qq^2}_9(s)=\rho^{\qq^2}_8(s)\, ,
\non
\rho^{\ss}_{10}(s)&=\frac{m_s\ss(s-m_Q^2)(9s^4-36m_Q^2s^3-m_Q^4s^2-m_Q^6s-m_Q^8)}{576\pi^4s^3}-\frac{5m_Q^4m_s\ss\log[m_Q^2/s]}{96\pi^4}\, ,
\non
\rho^{\GGa}_{10}(s)&=\frac{\GGb}{768\pi^6}\dab\Bigg[
-\frac{(1-\alpha-\beta)\f(s)(17\alpha m_s^2+17\beta m_Q^2-47\alpha\beta s)}{8\alpha^2\beta}
\non &+\frac{m_Q^2(1-\alpha-\beta)^2(4\alpha m_s^2+4\beta m_Q^2-9\alpha\beta s)}{2\alpha^3}-\frac{m_Q^2(1-\alpha-\beta)^3(\alpha m_s^2+\beta m_Q^2-3\alpha\beta s)}{3\alpha^3}
\non &-\frac{5\f(s)(\alpha m_s^2+\beta m_Q^2-3\alpha\beta s)}{8\alpha\beta}+\frac{(1-\alpha-\beta)^2\f(s)s}{\alpha}
\non &-\frac{(1-\alpha-\beta)^2\f(s)^2}{4\alpha^2\beta}\Bigg]\, ,
\\
\rho^{\qGqb}_{10}(s)&=\frac{5\qGqa}{384\pi^4}\dab\Bigg[13(\alpha m_s^2+\beta m_Q^2-2\alpha\beta s)\left(\frac{m_Q}{\alpha}+\frac{m_s}{\beta}\right)
\non & -\frac{(m_Q-m_s)(1-\alpha-\beta)(\alpha m_s^2+\beta m_Q^2-2\alpha\beta s)}{\alpha\beta}\Bigg]\, ,
\non
\rho^{\qq\qGqb}_{10}(s)&=\frac{5(7s^2+6m_Q^4)\qq\sGsa}{288\pi^2s^2}+\frac{5(s^2+m_Q^4)\ss\qGqa}{48\pi^2s^2}
\non &+\frac{5m_Qm_s\qq\qGqa}{288\pi^2}\int_0^1d\alpha
\Bigg[\frac{24m_s^2}{\alpha^2}\delta'\left(s-\tilde{m}^2_Q\right)+
\frac{1}{\alpha(1-\alpha)}\delta\left(s-\tilde{m}^2_Q\right)
\Bigg]\, . \label{SD2+S}
\end{align}
}

\item For the current $J_{3\mu}(x)$ with $J^P=0^-$
{\allowdisplaybreaks
\begin{align}
\nonumber
\rho^{pert}_{11}(s)&=-\frac{1}{1024\pi^6}\dab\frac{(1-\alpha-\beta)^2\f(s)^3(\alpha m_s^2+\beta m_Q^2+3\alpha\beta s)}{\alpha^3\beta^3}\, ,
\non
\rho^{\qq}_{11}(s)&=\frac{\qq}{16\pi^4}\dab\Bigg[\frac{m_Q(1-\alpha-\beta)\f(s)(\alpha m_s^2+\beta m_Q^2-2\alpha\beta s)}{\alpha^2\beta}
\non &+\frac{m_s(1-\alpha-\beta)\f(s)(\alpha m_s^2+\beta m_Q^2+\alpha\beta s)}{2\alpha\beta^2}\Bigg]\, ,
\non
\rho^{\ss}_{11}(s)&=\frac{m_s\ss(s-m_Q^2)(s^3+m_Q^2s^2+13m_Q^4s-3m_Q^6)}{1536\pi^4s^2}+\frac{m_Q^4m_s\ss\log[m_Q^2/s]}{128\pi^4}\, ,
\non
\rho^{\GGa}_{11}(s)&=-\frac{\GGb}{1024\pi^6}\dab\Bigg\{
\frac{(1-\alpha-\beta)^2(\alpha m_s^2+\beta m_Q^2)}{3}\left(\frac{m_Q^2}{\alpha^3}+\frac{m_s^2}{\beta^3}\right)
\non &-\frac{(1-\alpha-\beta)\f(s)}{2\alpha\beta}\left[\frac{\alpha m_s^2+\beta m_Q^2-3\alpha\beta s}{\alpha}-\frac{\alpha m_s^2+\beta m_Q^2+\alpha\beta s}{\beta}\right]\Bigg\}\, ,
\\
\rho^{\qGqb}_{11}(s)&=\frac{\qGqa}{64\pi^4}\dab\Bigg[\frac{m_s(1-\alpha-2\beta)(\alpha m_s^2+\beta m_Q^2)}{\beta^2}
\non &-\frac{m_Q(2\alpha m_s^2+2\beta m_Q^2-3\alpha\beta s)}{\alpha}\Bigg]\, ,
\non
\rho^{\qq^2}_{11}(s)&=-\frac{m_Qm_s\qq^2}{12\pi^2}\sqrt{\left(1-\frac{m_Q^2-m_s^2}{s}\right)^2-\frac{4m_s^2}{s}}
+\frac{m_Q\qq\ss(s-m_Q^2)(m_ss+m_Qs-m_Q^3)}{24\pi^2s^2}\, ,
\non
\rho^{\qq\qGqb}_{11}(s)&=-\frac{(s^2-3m_Q^4)(\ss\qGqa+\qq\sGsa)}{96\pi^2s^2}
\non &-\frac{m_Qm_s\qq\qGqa}{48\pi^2}\int_0^1d\alpha
\Bigg[\frac{2m_s^2}{\alpha^2}\delta'\left(s-\tilde{m}^2_Q\right)-
\frac{1}{\alpha}\delta\left(s-\tilde{m}^2_Q\right)
\Bigg]\, . \label{SD0-1}
\end{align}
}

\item For the current $J_{4\mu}(x)$ with $J^P=0^-$
{\allowdisplaybreaks
\begin{align}
\nonumber
\rho^{pert}_{12}(s)&=\rho^{pert}_{11}(s), \, \rho^{\ss}_{12}(s)=\rho^{\ss}_{11}(s)\, ,
\non
\rho^{\qq}_{12}(s)&=\frac{\qq}{16\pi^4}\dab\Bigg[\frac{m_Q(1-\alpha-\beta)\f(s)(\alpha m_s^2+\beta m_Q^2+\alpha\beta s)}{2\alpha^2\beta}
\non &+\frac{m_s(1-\alpha-\beta)\f(s)(\alpha m_s^2+\beta m_Q^2-2\alpha\beta s)}{\alpha\beta^2}\Bigg]\, ,
\non
\rho^{\GGa}_{12}(s)&=-\frac{\GGb}{1024\pi^6}\dab\Bigg\{
\frac{(1-\alpha-\beta)^2(\alpha m_s^2+\beta m_Q^2)}{3}\left(\frac{m_Q^2}{\alpha^3}+\frac{m_s^2}{\beta^3}\right)
\non &-\frac{(1-\alpha-\beta)\f(s)}{2\alpha\beta}\left[\frac{\alpha m_s^2+\beta m_Q^2-3\alpha\beta s}{\beta}-\frac{\alpha m_s^2+\beta m_Q^2+\alpha\beta s}{\alpha}\right]\Bigg\}\, ,
\\
\rho^{\qGqb}_{12}(s)&=\frac{\qGqa}{64\pi^4}\dab\Bigg[\frac{m_Q(1-2\alpha-\beta)(\alpha m_s^2+\beta m_Q^2)}{\alpha^2}
\non &-\frac{m_s(2\alpha m_s^2+2\beta m_Q^2-3\alpha\beta s)}{\beta}\Bigg]\, ,
\non
\rho^{\qq^2}_{12}(s)&=-\frac{m_Qm_s\qq^2}{12\pi^2}\sqrt{\left(1-\frac{m_Q^2-m_s^2}{s}\right)^2-\frac{4m_s^2}{s}}
-\frac{\qq\ss(s-m_Q^2)(2s^2-m_Qm_ss-2m_Q^2s+3m_Q^3m_s)}{48\pi^2s^2}\, ,
\non
\rho^{\qq\qGqb}_{12}(s)&=-\frac{\ss\qGqa+\qq\sGsa}{48\pi^2}
\non &-\frac{m_Qm_s\qq\qGqa}{48\pi^2}\int_0^1d\alpha
\Bigg[\frac{2m_s^2}{\alpha^2}\delta'\left(s-\tilde{m}^2_Q\right)-
\frac{1}{\alpha}\delta\left(s-\tilde{m}^2_Q\right)
\Bigg]\, . \label{SD0-2}
\end{align}
}

\end{itemize}


\end{document}